\documentclass[a4paper,11pt]{article}
\pdfoutput=1
\usepackage{jheppub}
\usepackage[T1]{fontenc}
\usepackage{amsfonts}
\usepackage{amsmath}
\usepackage{amssymb}
\usepackage{multirow}
\usepackage{graphicx}
\usepackage{mathtools}
\usepackage{comment}
\usepackage{caption}
\usepackage{subcaption}

\numberwithin{equation}{section}

\newcommand{\bea}{\begin{eqnarray}}
\newcommand{\eea}{\end{eqnarray}}

\newcommand{\be}{\begin{equation}}
\newcommand{\ee}{\end{equation}}

\begin{document}

\title{Quantum Corrections to $\eta/s$ from JT Gravity}

\author[a]{Sera Cremonini,}
\emailAdd{cremonini@lehigh.edu}
\affiliation[a]{Department of Physics, Lehigh University, Bethlehem, PA, 18015, USA}
\author[b,c]{Li Li,}
\emailAdd{liliphy@itp.ac.cn}
\affiliation[b]{Institute of Theoretical Physics, Chinese Academy of Sciences, Beijing 100190, China}
\affiliation[c]{School of Fundamental Physics and Mathematical Sciences, Hangzhou Institute for Advanced Study, UCAS, Hangzhou 310024, China}
\author[a]{Xiao-Long Liu,}
\emailAdd{xila25@lehigh.edu}

\author[d]{Jun Nian}
\emailAdd{nianjun@ucas.ac.cn}
\affiliation[d]{International Centre for Theoretical Physics Asia-Pacific (ICTP-AP), University of Chinese Academy of Sciences, 100190 Beijing, China}

\date{\today}

\abstract{
We revisit the computation of the shear viscosity  to entropy ratio $\eta/s$ at finite chemical potential in a holographic model that takes into account the quantum fluctuations in the IR region of near-extremal black branes. Such quantum corrections can be computed from JT gravity and generate non-trivial temperature dependence for $\eta/s$, which deviates from the universal $1/4\pi$ result. In the semi-classical regime, $\eta/s$
attains a minimum which is below the KSS bound, generated by the presence of the quantum effects. In the quantum regime at lower temperatures, $\eta/s$ increases and is well above the KSS bound.
We also compare the shear viscosity to the quantum-corrected absorption cross-section of near-extremal black holes, and find agreement.}

\maketitle

\section{Introduction}

Since the early days of the AdS/CFT correspodence, the techniques of holography have been applied to probe the dynamics of a wide spectrum of strongly correlated quantum phases of matter \cite{Casalderrey-Solana:2011dxg} -- from the QCD quark gluon plasma (QGP) to high temperature superconductors and strange metals, just to name a few.
A particularly useful result that has come out of this program is the 
universality \cite{Policastro:2001yc,Buchel:2003tz} of the shear viscosity $\eta$ to entropy density $s$ ratio, 
\begin{equation}
\label{universal}
\frac{\eta}{s} = \frac{\hbar}{4\pi k_B} \, ,
\end{equation}
which holds in strongly coupled gauge theories in the limit of an infinite number of colors, $N \rightarrow \infty$, and
infinite 't Hooft coupling, $\lambda \rightarrow \infty$. 
The simple result 
(\ref{universal}) is obtained by working with holographic theories that describe Einstein gravity coupled to an arbitrary matter sector, under the additional assumption that rotational invariance is preserved\footnote{For a recent discussion of $\eta/s$ in anisotropic theories we refer the reader to \cite{Baggioli:2023yvc,Zhao:2025gej}.}. From now on we take $\hbar=k_B=1$.

Apart from its elegance and universality, the importance of (\ref{universal}) is that it is remarkably close 
to the experimental range extracted from the QGP data at RHIC and at the LHC. 
Indeed, this led to the compelling KSS proposal \cite{Kovtun:2003wp,Kovtun:2004de} 
that the shear viscosity might obey 
a fundamental lower bound in nature, $ \frac{\eta}{s}  \geq \frac{1}{4\pi}$.
Attempts to further understand the structure of $\eta/s$ have led to the realization that 
the KSS bound can in fact be violated  -- either by relaxing symmetries or by introducing higher derivative curvature corrections to the low-energy gravitational action (see \emph{i.e.} \cite{Cremonini:2011iq} for a review). 
Apart from encoding deviations from the universal result (\ref{universal}), such holographic constructions are also useful in that they generate a temperature dependent $\eta/s$, which is of clear interest to the heavy ion community and the efforts to better 
understand QCD and the QGP (see \emph{e.g.} the recent review \cite{Rougemont:2023gfz}).

A natural question is whether the universal behavior (\ref{universal}) survives as the temperature approaches zero, $T\rightarrow 0$. The holographic computation of $\eta/s$ at precisely zero temperature and at finite charge density (using the
extremal Reissner-Nordstrom AdS$_4$ black hole) was done in \cite{Edalati:2009bi} and yielded once again the 
universal $1/4\pi$ result. 
However, extracting hydrodynamic quantities at strictly zero temperature is tricky, 
since $\omega$ and $k$ are no longer the smallest scales in the system.
Moreover, recent years have seen new insights into the thermodynamic behavior of black holes at very low temperatures. 
In particular, there has been remarkable progress 
in understanding the quantum
nature of the AdS$_2$ region that arises near the horizon of near-extremal black holes 
\cite{Almheiri:2014cka,Maldacena:2016upp,Jensen:2016pah,Almheiri:2016fws,Stanford:2017thb,Mertens:2017mtv,Mertens:2018fds,Lam:2018pvp,Mertens:2019tcm,Mertens:2022irh,Turiaci:2023wrh,Iliesiu:2020qvm,Sachdev:2019bjn}. Initial progress was achieved in the context
of Jackiw-Teitelboim (JT) gravity. 
The understanding of the nature
of the quantum fluctuations was later applied to higher-dimensional near-extremal black
holes and led to key modifications to their low-temperature thermodynamics \cite{Iliesiu:2020qvm,Heydeman:2020hhw,Boruch:2022tno,Iliesiu:2022onk}.

Given these recent developments, it is natural to ask how such quantum effects will affect transport coefficients such as $\eta/s$. 
Recall that in holography transport coefficients can be extracted in a number of complementary ways, \emph{e.g.} computing correlators of the stress energy tensor and using Kubo formulas, or from
linearized quasi-normal modes on black brane backgrounds, which in the hydrodynamic limit correspond to shear and sound modes of the dual field theory. 
The standard holographic dictionary instructs us to extract correlators from the \emph{boundary} behavior of fluctuating bulk fields, appropriately supplemented with boundary conditions at the horizon. 
However,  hydrodynamics is an effective description of the system at long wavelengths and small frequencies, and thus one expects it to be 
encoded in properties of the geometry and its fluctuations in the IR, \emph{i.e.} near the horizon.
This is precisely the part of the geometry where quantum corrections play a key role. 


Thus, our logic in this paper is the following. 
We compute quantum corrections to the IR retarded Green's function $\mathcal{G}_R(\omega,T)$, working in two different temperature regimes: one which we refer to as \emph{semiclassical}, for which the temperature scale dominates over the quantum scale, and one which we refer to as 
\emph{quantum}, in which the quantum effects dominate. 
We should note that, 
despite being in a low temperature regime,  we always work with $\omega\ll T$ to ensure that hydrodynamics is well defined.
We then relate $\mathcal{G}_R(\omega,T)$ to the retarded Green's function
$G_R(\omega,T)$ in the UV, using 
\cite{Faulkner:2009wj,Faulkner:2013bna,Faulkner:2010da}, and 
use Kubo's formula to extract the shear viscosity.

We should stress that -- unlike in the semiclassical regime, which is well-understood -- our computation of $\eta$ in the deep quantum regime should be interpreted with caution, since a new framework is likely needed to accurately capture hydrodynamics.  In this paper we will not address this challenge, but rather present our computations in both regimes, hoping that 
the results in the quantum regime 
might also provide guidance. 

Moreover, we should note that it's possible that certain 
instabilities will keep one from even accessing the extremal regime \cite{Buchel:2025ves,Buchel:2025jup,Gladden:2024ssb,Choi:2024xnv},  invalidating the need to incorporate the quantum fluctuations of the near-extremal brane. However, our goal in this paper is to investigate the quantum corrections of $\eta/s$ that arise 
under the assumption that the 
near-extremal AdS$_2$ throat is stable. Provided this is the case, our analysis characterizes the universal IR physics associated with JT gravity, which is independent of the detailed UV completion. Finally, for completeness one should examine when the quantum effects computed here 
start competing with those due to higher-derivative operators in the gravitational theory (see \cite{Buchel:2026xtn} for an efficient framework for extracting the behavior of \emph{e.g.} $\eta/s$ in such theories).
We will not examine these questions further in this paper.

To compute the quantum contributions to $\eta/s$, we will also need the quantum-corrected entropy density $s$ arising from the one-loop  corrections of JT gravity \cite{Turiaci:2023wrh,Iliesiu:2020qvm,Mertens:2022irh}. 
As is well known, the entropy formally turns negative\footnote{This could be avoided by working with the quenched free energy and taking into account additional wormholes (see Section \ref{Section4} for a brief discussion).} when the quantum correction, namely the logarithmic contribution from the JT mode, dominates at ultra-low temperatures. In this regime, the near-extremal entropy formula becomes unreliable -- the entropy becoming negative signals the breakdown of the semiclassical description.
Thus, one cannot take arbitrarily low temperatures when evaluating the quantum-corrected $\eta/s$.
We will also compare our results for $\eta$ to the quantum corrected absorption cross section of a massless scalar \cite{Emparan:2025sao} (see also \cite{Biggs:2025nzs}) and find agreement in the two temperature regimes of interest\footnote{The cross sections for photon and graviton scattering were also computed in \cite{Emparan:2025qqf}.}.

As we will see, the quantum corrections will generate a non-trivial temperature dependence for $\eta/s$, unlike the tree-level case for which 
$\eta/s=1/4\pi$.
Moreover, we will find that $\eta/s$ reaches a \emph{minimum} which violates the KSS bound in the semi-classical regime (and not in the deep quantum regime), 
suggesting that even a small amount of quantum corrections might lead to a lower bound.  
We will show that the minimum can be traced to a \emph{maximum} in the quantum corrected entropy, while the viscosity itself is monotonic.
We should also mention that 
recent studies \cite{PandoZayas:2025snm,Nian:2025oei,Gouteraux:2025exs,Kanargias:2025vul} have investigated quantum corrections to hydrodynamic transport coefficients arising from the JT mode at low temperatures\footnote{The earlier work \cite{Jensen:2016pah} also attempted to rewrite a generic theory of gravity near the AdS$_2$ throat as a novel hydrodynamics coupled to the correlation functions of a conformal quantum mechanics.}. As we show explicitly in Section 
\ref{Section4}, 
our results for $\eta/s$ are in agreement with  those of  \cite{PandoZayas:2025snm,Kanargias:2025vul}. The analysis of 
\cite{Gouteraux:2025exs} considered a different conformal dimension from the one studied here and in \cite{PandoZayas:2025snm,Kanargias:2025vul}, and therefore yields different results.

The structure of the paper is as follows.
Section \ref{Section2} reviews the standard holographic computation of $\eta/s$ in Einstein gravity coupled to matter, following closely the analysis of \cite{Benincasa:2006fu} for convenience.
Section \ref{Section3} outlines the connection between the retarded Green's functions of the IR and UV regions.
In Section \ref{Section4} we compute the effects of quantum corrections on the IR Green's function and use them to extract the quantum-corrected UV Green's function.
We apply the latter to compute $\eta/s$ and study its temperature dependence in various regimes. Finally, Section \ref{Section5} compares the shear viscosity to the quantum-corrected cross section of \cite{Emparan:2025sao}, finding agreement.
We conclude with open questions and subtleties of the computations.

\section{Holographic $\eta/s$ at finite chemical potential}
\label{Section2}

Before we examine the effects of quantum corrections on $\eta/s$,  we briefly review the by now standard tree-level computation that yields the simple universal relation between shear viscosity and entropy density, 
\begin{equation} 
\label{etaprops}
    \frac{\eta}{s}= \frac{1}{4\pi} \, .
\end{equation}
Holographic computations of $\eta/s$ in the presence of a chemical potential in theories with an Einstein dual\footnote{For higher derivative effects at finite chemical potential see \cite{Myers:2009ij,Cremonini:2009sy}.} 
were carried out in a number of early papers \cite{Mas:2006dy,Son:2006em,Saremi:2006ep,Maeda:2006by}, which confirmed  (\ref{etaprops}).
We remind the reader that one of the ways to compute the shear viscosity relies on using Kubo's formula, 
which relates $\eta$ to the low frequency and zero momentum limit of
the retarded Green’s function of the CFT stress tensor,
\begin{eqnarray}
\label{etastress}
\eta &=& - \lim_{\omega \rightarrow 0} \frac{1}{\omega} \,
    \text{Im}\, G^R_{xy,xy}(\omega,0) \, , \nonumber \\
    G^R_{xy,xy}(\omega,0)  &=& - i  \frac{1}{\omega} \,
    \int dt d\vec{x} \, e^{i \omega t}\theta(t)\,  \langle [T_{xy}(t,\vec{x}),T_{xy}(0,0)]\rangle \, .
\end{eqnarray} 
where only two spatial dimensions $(x,y)$ are considered for simplicity. 
Using the holographic prescription of \cite{Son:2002sd,Policastro:2002se},
the retarded Green’s function can be extracted 
from the effective action for the metric perturbation dual to the shear mode, as we will sketch below.

To show how to obtain (\ref{etaprops}) using Kubo's formula, we find it particularly useful to 
follow the holographic analysis of \cite{Benincasa:2006fu}, which  examined a broad class of models 
\begin{equation}
\label{GeneralAction}
    S= \frac{1}{2 \kappa^2} \int d^D x \sqrt{-g} \left[R - \mathcal{K}_{\alpha \beta}(\phi) \partial_\mu \phi^\alpha \partial^\mu \phi^\beta 
    - \mathcal{V}(\phi) - \tau_{ab}{(\phi)} F_{\mu \nu}^{a} F^{\mu \nu \, b}   \right] ,
\end{equation}
involving an arbitrary number of scalar fields $\phi^\alpha$ and vectors $A_\mu^{a}$, with fluxes given by
$F_{\mu \nu}^{a} = \partial_\mu A_\nu^{a}- \partial_\nu A_\mu^{a}$.
The theory admits a black $D$-dimensional brane solution, electrically charged under the vector fields $A_\mu^a$.
The advantage of the analysis of \cite{Benincasa:2006fu} is that uses a general background geometry (preserving $SO(D)$ invariance), and thus can be easily adopted and applied to the case we are after.

Since we are interested in working with charged black brane solutions to 
Einstein-Maxwell theory in four dimensions, we turn off all the scalar fields in (\ref{GeneralAction}) and keep only one $U(1)$ gauge field $A_\mu$, with $F_{\mu \nu} = \partial_{\mu} A_{\nu}- \partial_\nu A_\mu$.
We write the action as follows,
\begin{equation}
\label{EinsteinMaxwell}
    S= \frac{1}{2 \kappa^2} \int d^4 x \sqrt{-g} \left[R +\frac{6}{L^2} - \frac{L^2}{g_F^2} F^2  \right] ,
\end{equation}
where $L$ denotes the radius of $AdS_4$ and $g_F^2$ the effective (dimensionless) gauge coupling.
Following the setup of \cite{Benincasa:2006fu}, we take the background metric and gauge field to be of the form
\begin{eqnarray}\label{eq:general}
&&    ds^2 = -c_1^2(r) dt^2 + c_2^2(r) (dx^2 + dy^2) + c_3^2(r) dr^2 \, .\nonumber \\
  && A_\mu = \delta_\mu^t \Phi(r) \, .
\end{eqnarray}
While for now it's convenient to keep 
the metric components $c_i(r)$ general, we are ultimately interested in a black brane solution in $AdS_4$, with $r_h$ denoting the black brane horizon and $r_b$ the AdS boundary.
To compute the shear viscosity it suffices to add a perturbation $\delta g_{xy}$ to the metric (\ref{eq:general}). 
Working with the 
following combination of the shear perturbation, $\psi = \frac{1}{2} \, c_2^{-2} \, \delta g_{xy}$, 
one can show that the quadratic effective action for the shear mode reduces to a boundary term\footnote{We refer the reader to \cite{Benincasa:2006fu} for details on the role of the Gibbons-Hawking boundary term, counterterms and the regularized bulk action for the perturbation.} of the form
\begin{equation}
    S_{\text{boundary}} [\psi_b] = \int \frac{d^3k}{(2\pi)^3} \psi^b(-\omega) \mathcal{F}(\omega,r) \psi^b(\omega) \bigg|_{r_h}^{r_b} \, .
\end{equation}
The retarded Green's function is then given by the boundary limit of the \emph{flux} $\mathcal{F}(\omega,r)$,
\begin{equation}
    G^R_{xy,xy}(\omega,\vec{k}=0)= -  \lim_{r \rightarrow r_b} 2 \, \mathcal{F}(\omega,r) \, .
\end{equation}
In terms of the background metric in (\ref{eq:general}), the regularized boundary term becomes 
\cite{Benincasa:2006fu} 
\begin{equation}
\label{eq:bound2}
    S_{\text{boundary}} [\psi]^{reg} = \frac{1}{2\kappa^2}\int_{\partial \mathcal{M}} dt \, d^3 x \left(- \frac{c_1 \, c_2^2}{2 c_3} \right) \psi \partial_r \psi \, .
\end{equation}

To obtain an explicit expression for the flux, we need to make use of the actual solution for the shear mode $\psi$. 
From the bulk quadratic effective action for $\psi =e^{- i \omega t} \psi_\omega(r)$ one obtains the following 
equation of motion \cite{Benincasa:2006fu},  
\begin{equation}
\label{sheareom}
    \frac{\omega^2}{c_1^2} \, \psi_\omega + 
    \frac{1}{c_3^2} \, \psi_\omega^{\prime\prime} + \frac{1}{c_3^2} 
    \left(\ln \frac{c_1 \, c_2^2}{c_3^2}\right)^\prime   \psi_\omega^{\prime}=0 \, ,
\end{equation}
where primes denote radial derivatives.
In the hydrodynamic regime (low-frequency approximation) one can neglect the $\omega^2$ term in the equation of motion, and easily obtain the following general solution, which is valid at an arbitrary location $r$,
\begin{equation}
\label{generalsol}
    \psi_\omega (r) = \mathcal{A}_1 (\omega) +
    \mathcal{A}_2(\omega) \, \int^{\infty}_r
    \frac{c_3(\rho)}{c_1(\rho) \, c_2^2(\rho)} \, d\rho \, .
\end{equation}
Expanding the solution near asymptotic infinity, 
where $c_1 (r) = c_2 (r) =\frac{1}{c_3 (r)} =\frac{r}{L}$, 
yields\footnote{The four-dimensional retarded Green's function read off from the asymptotic expansion is 
$G_R \sim \frac{\mathcal{A}_2(\omega) }{\mathcal{A}_1(\omega) }$.} 
\begin{equation}
   \psi_\omega (r) \sim \mathcal{A}_1 (\omega) - \mathcal{A}_2(\omega) \frac{L^4}{r^3} \, ,
\end{equation}
where
$\mathcal{A}_1 (\omega)$ can be fixed by imposing that the bulk perturbation approaches a constant mode at the boundary.
Choosing a convenient normalization, we take $\mathcal{A}_1 (\omega)=1$.
On the other hand, by examining \cite{Benincasa:2006fu} the near-horizon region of the geometry one can relate 
$\mathcal{A}_2(\omega)$ to the black hole entropy, 
\begin{equation}
    \mathcal{A}_2(\omega)= \frac{\kappa^2}{2\pi} \, i\,\omega \,  s\, .
\end{equation}
Finally, using \eqref{generalsol}
to compute
\begin{equation}
\label{deriv}
    \partial _r \psi_\omega (r) = 
     - \mathcal{A}_2(\omega) 
    \frac{c_3}{c_1 \, c_2^2}  \, ,
\end{equation}
and  (\ref{eq:bound2}) to extract the flux,
one finds, working to 
leading order in $\omega$, 
\begin{equation}\label{eq:ga}
G_R(\omega) = -\frac{\mathcal{A}_2(\omega)}{2\kappa^2} \, .
\end{equation}
Combining these ingredients and using Kubo's formula, one recovers the universal result
\begin{eqnarray}
\eta &=& - \lim_{\omega \rightarrow 0} \frac{1}{\omega} \,
    \text{Im}\, G_R(\omega,0) 
    =  \frac{1}{4\pi} \, s \, . 
\end{eqnarray}

\section{UV and IR retarded Green's functions}\label{Section3}

Next, we want to show how the retarded Green's function of the IR AdS$_2$ geometry can be related to that extracted from the UV.  
Einstein-Maxwell theory \eqref{EinsteinMaxwell} admits as a solution 
the AdS$_4$ charged black brane geometry described by
\begin{align}
ds^2 & = \frac{r^2}{L^2}\left[-h(r)dt^2 + dx^2+dy^2\right] + \frac{L^2}{r^2}\frac{dr^2}{h(r)}\, ,\label{eq:metricads4}\\
  A_t & = \mu \left(1-\frac{r_h}{r}\right)\, ,\quad 
  h(r)=1+\frac{Q^2}{r^4}-\frac{M}{r^3} \, ,
  \label{eq:gaugeAdS4}
\end{align}
where $\mu$ denotes the chemical potential and $r_h$ the outer horizon radius
determined by the largest positive root of the blackening factor,
\begin{equation}
h(r_h)=0\quad\rightarrow\quad M=r_h^3+\frac{Q^2}{r_h} \, .
\end{equation}
The corresponding entropy density, temperature and chemical potential 
are then
\begin{equation}
s = \frac{2\pi}{\kappa^2}\left(\frac{r_h}{L}\right)^2,\quad  T = \frac{3r_h}{4\pi L^2} \left(1-\frac{Q^2}{3r_h^4}\right),\quad \mu = \frac{g_F\, Q}{L^2r_h} \, .
\end{equation}
We are going to work in an ensemble with fixed charge, and take 
$Q=\sqrt{3}r^2_0$, where $r_0$ denotes the extremal horizon radius. 
We can then expand $r_h$ and $s$ in powers of $T$ in a low temperature expansion,
\begin{equation}
r_h\simeq r_0+\frac{\pi L^2}{3}T+\frac{\pi ^2 L^4}{6 r_0}T^2,\quad 
s 
\simeq \frac{2\pi}{\kappa^2}\left(\frac{r_0}{L}\right)^2\left(1+\frac{2 \pi  L^2}{3 r_0}T+\frac{4 \pi ^2 L^4 }{9 r_0^2}T^2\right) \, .
\end{equation}
Next, we examine the near horizon AdS$_2$ region of the charged brane. Following \cite{Faulkner:2009wj}, we consider the following scaling limit,
\begin{align}\label{eq:ads2limit}
r-r_0 \to \lambda\frac{L^2_2}{\zeta},\qquad r_h-r_0 \to \lambda\frac{L^2_2}{\zeta_0},\qquad t\to\lambda^{-1}t,
\end{align}
further taking $\lambda\to 0$ with $\zeta$, $\zeta_0$, and $\tau$ finite. 
The parameter $L_2$ denotes the AdS$_2$ radius. The scaling \eqref{eq:ads2limit} defines a new variable $\zeta$ and a new length scale $\zeta_0$. In this near-horizon limit, the metric \eqref{eq:metricads4} becomes that of AdS$_2\times\mathbb{R}^{2}$, 
\begin{equation}\label{eq:metricads2}
ds^2=\frac{L^2_2}{\zeta^2} \left[- \left(1-\frac{\zeta^2}{\zeta^2_0}\right) dt^2 +\frac{d\zeta^2}{1-\frac{\zeta^2}{\zeta^2_0}} \right] + \frac{r^2_h}{L^2}d\vec{x}\,^2\, ,
\end{equation}
and the gauge field \eqref{eq:gaugeAdS4} becomes
\be
  A_{t}=\frac{g_F}{2\sqrt{3}} \left(\frac{1}{\zeta}-\frac{1}{\zeta_0}\right)\, .
\ee
Consequently, the Hawking temperature is $T=\frac{1}{2\pi\zeta_0}$, and the AdS$_2$ radius is $L_2=L/\sqrt{6}$. 

Next, consider a charged scalar field in AdS$_2$ of charge $q$ and mass $m$,  
dual to an operator $\mathcal{O}$ in the boundary CFT$_1$ of charge $q$ and dimension $\ell$.
Solving the equation of motion for the scalar in the background AdS$_2$ geometry \eqref{eq:metricads2}, one finds \cite{Faulkner:2009wj}, up to a constant,
\begin{equation}\label{eq:wave1}
\phi(r)=1+\mathcal{G}_R(\omega,T)L^2_2(r-r_0)^{-1} \, ,
\end{equation}
where $\mathcal{G}_R$ denotes the retarded IR Green's function for the AdS$_2$/CFT$_1$ theory,
\begin{equation}\label{eq:treelevelt}
\mathcal{G}_R(\omega,T)=(4\pi T)^{2\ell-1}\frac{\Gamma(1-2\ell)\Gamma(\ell-\frac{i\omega}{2\pi T}+iqe_d)\Gamma(\ell-iqe_d)}{\Gamma(2\ell-1)\Gamma(1-\ell-\frac{i\omega}{2\pi T}+iqe_d)\Gamma(1-\ell-iqe_d)}  \, .
\end{equation}
We have introduced  $e_d \equiv g_F / 2\sqrt{3}$ and 
\begin{equation}
\ell = \sqrt{m^2L^2_2-q^2e^2_d+\frac{1}{4}}+\frac{1}{2} \, .
\end{equation}
Since we are interested in the shear viscosity, which corresponds to a neutral (massless) mode, we take a vanishing charge ($q=0$), for which\footnote{Note that \eqref{eq:ads2tree} can be written as $
\mathcal{G}_R(\omega,T) = (4\pi T)^{2\ell-1}\frac{\Gamma(1-2\ell)\Gamma(\ell-\frac{i\omega}{2\pi T})\Gamma(\ell)}{\Gamma(2\ell-1)\Gamma(1-\ell-\frac{i\omega}{2\pi T})\Gamma(1-\ell)}$ after making use of Gamma function identities. See \emph{e.g.} \cite{Faulkner:2009wj,Edalati:2009bi}.}
\begin{equation}\label{eq:ads2tree}
\boxed{
\mathcal{G}_R(\omega,T) = -(\pi T)^{2\ell-1}\frac{\Gamma(\ell-\frac{i\omega}{2\pi T})}{\Gamma(1-\ell-\frac{i\omega}{2\pi T})}\frac{\Gamma(\frac{3}{2}-\ell)}{\Gamma(\frac{1}{2}+\ell)} \,,}
\end{equation}
and 
\begin{equation}
\ell = \sqrt{m^2L^2_2+\frac{1}{4}}+\frac{1}{2} \, .
\end{equation}
For now we keep the scalar mass arbitrary -- we will take the massless limit appropriate for the shear mode later on.

Next, we would like to relate the AdS$_2$/CFT$_1$ retarded Green's function 
\eqref{eq:treelevelt} to that of AdS$_4$/CFT$_3$. 
To put the AdS$_2$ wavefunction into a form that is particularly 
convenient for a direct comparison with the discussion of Section \ref{Section2}, we 
perform the inverse coordinate transformation to \eqref{eq:ads2limit}, so that 
the AdS$_2$ metric \eqref{eq:metricads2} can be written in the form
\begin{equation}
ds^2=-\frac{(r-r_0)^2-(r_h-r_0)^2}{L^2_2}dt^2+\frac{L^2_2}{(r-r_0)^2-(r_h-r_0)^2}dr^2+\frac{r^2_h}{L^2}d\vec{x}^2 \,.
\end{equation}
Comparing this to the near horizon behavior of the general metric~\eqref{eq:general}, we read off
\begin{equation}
c_1\to\frac{1}{L_2}\sqrt{(r-r_0)^2-(r_h-r_0)^2},\quad c_2\to\frac{r_h}{L},\quad c_3\to L_2\frac{1}{\sqrt{(r-r_0)^2-(r_h-r_0)^2}} \, .
\end{equation}
Plugging these  back into the general wavefunction solution given in~\eqref{generalsol} with $\mathcal{A}_1=1$ and taking the boundary limit of the near horizon AdS$_2$ region, we obtain 
\begin{equation}
\psi(r)\approx1+\mathcal{A}_2(\omega)\frac{ L^4}{6r^2_h}(r-r_0)^{-1} \, .
\end{equation}
Comparing with ~\eqref{eq:wave1}, one can read off
\begin{equation}
\mathcal{A}_2(\omega)=\frac{r_h^2}{L^2}\mathcal{G}(\omega,T) \, .
\end{equation}
Finally, plugging this back into ~\eqref{eq:ga}, we have the following relationship between and UV and IR retarded Green's functions,
\begin{equation}\label{eq:2d4d}
\boxed{
G_R(\omega,T) = -\frac{\mathcal{A}_2(\omega)}{2\kappa^2}
=-\frac{1}{2\kappa^2}\left(\frac{r_h}{L}\right)^2 \mathcal{G}_R(\omega,T) \, .}
\end{equation}
Note that this expression is a special case of the more general result of \cite{Faulkner:2009wj,Faulkner:2013bna,Faulkner:2010da}.

The tree level  shear viscosity corresponds to the massless limit of the case 
examined above, \emph{i.e.} $\ell=1$, which gives 
\begin{equation}
\eta = -\lim_{\omega\to0}\frac{1}{\omega}{\rm Im}[ G_R(\omega,T)]= \frac{1}{2\kappa^2}\left(\frac{r_h}{L}\right)^2T^{2\ell-2} \bigg|_{\ell=1} \approx \frac{1}{2\kappa^2}\left(\frac{r_0}{L}\right)^2\left(1+\frac{2 \pi  L^2}{3 r_0}T\right) \, .
\end{equation}
Finally, combining this expression with the tree level entropy, one finds 
\begin{equation}
\frac{\eta}{s} = \frac{\frac{1}{2\kappa^2}\left(\frac{r_0}{L}\right)^2\left(1+\frac{2 \pi  L^2}{3 r_0}T\right)}{\frac{2\pi}{\kappa^2}\left(\frac{r_0}{L}\right)^2\left(1+\frac{2 \pi  L^2}{3 r_0}T\right)} = \frac{1}{4\pi} \, .\label{eq:shear0}
\end{equation}

\section{Quantum corrected Green's function and $\eta/s$}\label{Section4}

This section contains the main results of our work. After motivating in Section \ref{Section41} the origin and role of quantum corrections arising from the deep IR geometry of the black branes we are interested in, 
we move on in Section \ref{Section42} to the computation of the quantum-corrected retarded Green's function, examining different temperature regimes. Technical details of the computation are relegated to the Appendix.
Finally, in Section \ref{Section43} we compute the quantum-corrected shear viscosity and -- making use of the quantum-corrected entropy -- study the temperature dependence of $\eta/s$ both in the semiclassical regime (where temperature dominates over the scale of quantum corrections) and in the 
opposite quantum regime where the temperature is subleading. 

\subsection{Quantum corrections from JT gravity}
\label{Section41}

The limiting near-horizon procedure introduced in Eq.~\eqref{eq:ads2limit} leads to certain zero modes in the gravitational path integral. By considering these zero modes in near-extremal black hole solutions with an AdS$_2$ throat, the thermodynamics description will be modified as explained in \cite{Iliesiu:2020qvm, Heydeman:2020hhw, Boruch:2022tno}, and \cite{Kapec:2023ruw, Rakic:2023vhv} for rotating black holes, including \cite{Maulik:2024dwq}, which established the result in various dimensions and extended it to asymptotically AdS black holes. 

The insight that temperature effectively functions as a coupling constant -- rendering the high-temperature regime classical while the low-temperature regime becomes quantum and strongly coupled -- was first established in the context of two-dimensional JT gravity \cite{Almheiri:2014cka, Jensen:2016pah, Maldacena:2016upp} (see \cite{Mertens:2022irh, Turiaci:2023wrh} for reviews). The current understanding is that any higher-dimensional gravitational theory containing near-extremal solutions with a near horizon AdS$_2$ throat will admit such zero modes and, as a result, the low-temperature thermodynamics and possibly other dynamical properties will be accordingly corrected.  

We briefly recall certain aspects of a specific dilaton gravity theory in AdS$_2$, JT gravity, which exhibits the same pattern of low-energy symmetry breaking as the (0+1)-dimensional quantum mechanical Sachdev–Ye–Kitaev (SYK) model \cite{Sachdev:1992fk,Kitaev,Maldacena:2016hyu}. With suitable boundary conditions, fluctuations of the AdS$_2$ background are governed by an effective (0+1)-dimensional Schwarzian action \cite{Maldacena:2016upp},
\begin{equation}
S[f]=-C\int^{\beta}_0 d\tau \, \{f,\tau\},\quad \text{with } \{f,\tau\} \equiv \frac{f'''}{f'} - \frac{3}{2} \left(\frac{f''}{f'}\right)^2\, ,
\end{equation}
where primes denote derivatives with respect to the Euclidean time $\tau$, $\beta$ is the inverse of temperature, and $C=\frac{r_0L^2_2}{G_N}$ is the Schwarzian coupling constant with length dimension. One can reach the semi-classical regime by taking $C/\beta$ to be large and reach the quantum regime by taking $C/\beta$ to be small. The function $f(\tau)$ denotes boundary time reparameterization and should not be confused with the blackening function of Eq.~\eqref{eq:gaugeAdS4}. One can introduce time fluctuations around the classical configuration by taking $f(\tau)=\tau+\epsilon(\tau)$, where $\epsilon(\tau)$ denotes  boundary time fluctuations. From the Schwarzian action, one can compute the correlation function of the fluctuation $\epsilon$ as \cite{Maldacena:2016upp, Maldacena:2016hyu, Mertens:2022irh},
\be
  \langle\epsilon(\tau)\, \epsilon(0)\rangle = \frac{1}{2\pi C} \left(\frac{\beta}{2\pi}\right)^3 \left[1-\frac{1}{2}\left(\frac{2\pi\tau}{\beta}-\pi\right)^2+\frac{\pi^2}{6} + \frac{5}{2}{\rm cos}\frac{2\pi\tau}{\beta} + \left(\tau-\pi\right)\, {\rm sin}\frac{2\pi\tau}{\beta}\right]\, .
\ee
For the Reissner-Nordstr\"om black brane \eqref{eq:metricads4} considered in this paper, in principle quantum fluctuations from the gauge field sector should also be taken into account. In this manuscript, however, we are interested in the shear mode, which corresponds to a massless and neutral scalar field that does not interact with the gauge field. Thus, such fluctuations can be ignored. Nevertheless, we discuss this point here for completeness and generality. The effective action for both gravity and gauge fluctuations is given by \cite{Sachdev:2019bjn, Mertens:2019tcm,Davison:2016ngz}:
\be\label{eq:effectiveaction}
  S_{eff}[f,\Lambda] = - C\int^{\beta}_{0} d\tau\, \Bigg\{{\rm tan}\frac{\pi}{\beta}f(\tau),\tau\Bigg\} -\frac{K}{2}\int^{\beta}_0 d\tau\, \left[\Lambda'(\tau)+i\frac{2\pi\mathcal{E}}{\beta} f'(\tau)\right]^2\, ,
\ee
where $\mathcal{E}=\frac{L^2_2Q_0}{4\pi r^2_0}$, with $Q_0$ the extremal black brane charge, $\Lambda(\tau) \equiv \int^{\infty}_{r_0}A_r(r,\tau)\, dr$ denotes the gauge fluctuation, and the coupling constant $K$ is the compressibility of the boundary quantum system, whose value has been discussed in \cite{Davison:2016ngz, Mertens:2019tcm} and is of the same order as $C$. As shown in Eq.~\eqref{eq:metricads2}, the near-horizon geometry of the Reissner-Nordstr\"om black brane is AdS$_2\times \mathbb{R}^2$. In principle, we should also consider the quantum fluctuations of $\mathbb{R}^2$. These can be handled with by imposing periodic boundary conditions on $\mathbb{R}^2$, which essentially replaces $\mathbb{R}^2$ by a torus $\mathbb{T}^2$ with $U(1) \times U(1)$ isometry, and then performing a Kaluza-Klein reduction. As in the AdS$_4$ black hole case \cite{Iliesiu:2020qvm}, the resulting JT gravity effective action from the AdS$_4$ black brane includes an additional contribution from the $U(1) \times U(1)$ gauge fields. A similar analysis to that of \cite{Iliesiu:2020qvm} showed that the energy scale $M_{U(1)\times U(1)}$, where the $U(1)\times U(1)$ gauge fluctuations start to dominate, satisfies
\be\label{eq:mc}
  M_{U(1)\times U(1)} \ll C^{-1} \simeq K^{-1}\, .
\ee
In this manuscript, we focus on the temperature regime $T \gg M_{U(1)\times U(1)}$. Hence, in \eqref{eq:effectiveaction}, we neglect the quantum fluctuations from $\mathbb{T}^2$ and consider it to be a classical background.

\subsection{Quantum corrected retarded Green's function}
\label{Section42}

We are now ready to compute the quantum corrected AdS$_2$/CFT$_1$ scalar Green's function $\mathcal{G}(\omega,T)$. This will then be used to obtain the AdS$_4$/CFT$_3$ Green's function $G(\omega,T)$, and in turn $\eta/s$, by applying the standard relation \eqref{eq:2d4d}.
As we will see, we will work in a hydrodynamic regime and examine two distinct temperature ranges, 
high and low as compared to the scale governing the quantum effects. We relegate comments about subtleties arising at low temperature to the Conclusions.

We start by performing a Fourier transformation of the tree-level Green's function \eqref{eq:treelevelt} and a Wick rotation $t\to-i\tau$ which yields, up to a constant\footnote{The constant is given by $B=2^{2\ell-1}\frac{\Gamma(1-2\ell)\Gamma(\ell-iqe_d)}{\Gamma(2\ell-1)\Gamma(1-\ell-iqe_d)}$ and, for $q=0$, by $B=-2^{-2\ell+1}\frac{\Gamma(\frac{3}{2}-\ell)}{\Gamma(\frac{1}{2}+\ell)}$. For simplicity, we temporarily omit $B$. We will  reinstate it in the last step of the  calculation.},
\begin{equation}\label{eq:saddle}
\mathcal{G}(\tau_1,\tau_2)= e^{-\frac{2\pi e_dq}{\beta}(\tau_1-\tau_2)}\left(\frac{1}{\frac{\beta}{\pi}\, {\rm sin}\frac{\pi}{\beta}\, |\tau_1-\tau_2|}\right)^{2\ell}\, .
\end{equation}
We note that this is also the tree level Green's function for the complex SYK model \cite{Davison:2016ngz}.
By introducing quantum fluctuations, $\mathcal{G}(\tau_1,\tau_2)$ becomes
\be
  \mathcal{G}'(\tau_1,\tau_2) = e^{-\frac{2\pi e_dq}{\beta}(f(\tau_1)-f(\tau_2))}\, e^{i(\Lambda(\tau_1)-
  \Lambda(\tau_2))} \left(\frac{\sqrt{f'(\tau_1)f'(\tau_2)}}{\frac{\beta}{\pi}\, {\rm sin}\frac{\pi}{\beta}\, |f(\tau_1)-f(\tau_2)|}\right)^{2\ell}\, .\label{eq:arbitrary}
\ee
Summing all the fluctuations, we obtain the quantum-corrected Green's function 
$\langle \mathcal{G}(\tau_1,\tau_2)\rangle$
as a path integral, with the effective action given in Eq.~\eqref{eq:effectiveaction},
\begin{equation}\label{eq:path}
\langle \mathcal{G}(\tau_1,\tau_2)\rangle=\int[\mathcal{D}f][\mathcal{D}\Lambda]\, e^{-S_{eff}[f,\Lambda]}\, \mathcal{G}'(\tau_1,\tau_2)\, .
\end{equation}
It's convenient to introduce $\tilde{\Lambda}(\tau)=\Lambda(\tau)-i\mu f(\tau)$, which leads to a decoupling of the  $f$-dependent and the $\Lambda$-dependent factors. The path integral \eqref{eq:path} then results in 
\begin{align}
\langle \mathcal{G} (\tau_1,\tau_2)\rangle&=\langle e^{i(\tilde{\Lambda}(\tau_1)-\tilde{\Lambda}(\tau_2))}\rangle\times \Bigg\langle \left(\frac{\sqrt{f'(\tau_1)f'(\tau_2)}}{\frac{\beta}{\pi}\, {\rm sin}\frac{\pi}{\beta}\, |f(\tau_1)-f(\tau_2)|}\right)^{2\ell}\Bigg\rangle\nonumber\\
& = \langle \mathcal{G}_f(\tau_1,\tau_2)\rangle\times \langle \mathcal{G}_{\tilde{\Lambda}}(\tau_1,\tau_2)\rangle\, ,
\end{align}
where the ${\rm SL}(2,\mathbb{R})$ contribution $\langle\mathcal{G}_f(\tau_1,\tau_2)\rangle$ and the ${\rm U}(1)$ contribution $\langle \mathcal{G}_{\tilde{\Lambda}}(\tau_1,\tau_2)\rangle$ have been solved in \cite{Mertens:2017mtv} and \cite{Mertens:2019tcm}, respectively. They have the following explicit expressions, up to a normalization\footnote{We choose a normalization for the quantum-corrected Green's function to ensure that in the classical limit we can recover the universal KSS result.},
\begin{align}
\langle \mathcal{G}_{f}(\tau)\rangle  &= \frac{1}{Z(\beta)}\frac{e^{S_0}}{\pi^2(2C)^{2\ell}} \int d\zeta(k_1)\, d\zeta(k_2)\, e^{-\mid\tau\mid \frac{k^2_1}{2C} - (\beta-\mid\tau\mid)\frac{k^2_2}{2C}}\, \frac{\Gamma(\ell\pm i(k_1\pm k_2))}{\Gamma(2\ell)}\, ,\label{eq:correction1}\\
\langle \mathcal{G}_{\tilde{\Lambda}}(\tau)\rangle & = e^{-\frac{\tau(\beta-\tau)}{2K\beta}}e^{\mu\tau}\frac{\theta_3 \left(i\frac{2\pi K}{\beta},-\frac{2\pi K}{\beta}\frac{\mu\beta}{2\pi}-\frac{\tau}{\beta}\right)}{\theta_3 \left(i\frac{2\pi K}{\beta},-\frac{2\pi K}{\beta}\frac{\mu\beta}{2\pi}\right)} \label{eq:correction2}\, ,
\end{align}
where $\tau_1=0,\;\tau=\tau_2-\tau_1$, $Z(\beta)=e^{S_0+\frac{2\pi^2 C}{\beta}+\frac{3}{2}{\rm log}\frac{2\pi C}{\beta}}$ is the JT gravity partition function with one-loop corrections, $S_0$ is the classical extremal entropy and $\theta_3$ is the theta function.

Since in this paper we are interested in the shear viscosity, which corresponds to a scalar field with no charge, we only need to consider quantum correction from the gravity fluctuations. In the following, we will drop the subscript $f$ in $\langle\mathcal{G}_f(\tau)\rangle$ for simplicity. Thus, we consider \eqref{eq:correction1}, which takes the form
\begin{align}
\langle \mathcal{G}(t)\rangle&
 = \frac{e^{S_0}}{Z(\beta)}\frac{4}{\pi^2(2C)^{2\ell}}\int k_1 dk_1\, k_2 dk_2\, {\rm sinh}(2\pi k_1)\, {\rm sinh}(2\pi k_2) e^{-it \frac{k^2_1}{2C}-(\beta-it)\frac{k^2_2}{2C}} \frac{1}{\Gamma(2\ell)}\nonumber\\
 & \quad\times \Gamma(\ell+i(k_1+k_2)) \Gamma(\ell-i(k_1+k_2)) \Gamma(\ell+i(k_1-k_2)) \Gamma(\ell-i(k_1-k_2))\, ,\label{eq:g1}
\end{align}
where we used $d\zeta(k)=dk^2{\rm sinh}(2\pi k)$, assumed $\tau > 0$ and  performed the Wick rotation $\tau\to it$. 
Following \cite{Lam:2018pvp}, we express $k_{1, 2}$ as
\begin{equation}
k^2_1=2CE_1=2C(M+\omega),\qquad k^2_2=2CE_2=2CM\, ,
\end{equation}
where $\omega$ denotes the energy of the scalar field under consideration, which describes a black hole perturbation, $E_2$ and $E_1$ represent the black hole energies before and after the perturbation, and $M$ is the black hole mass.

In what follows, we will compute the Wightman Green's function $\langle\mathcal{G}(\omega)\rangle$ in two distinct temperature regimes, and use the KMS condition 
\begin{equation}
{\rm Im}\;\langle\mathcal{G}_R(\omega,\beta)\rangle = \frac{1}{2}\left(1-e^{-\beta\omega}\right)\langle\mathcal{G}(\omega)\rangle\,,
\end{equation}\label{eq:kms}
to obtain the corresponding retarded Green's functions. 
First, we will examine the \emph{semi-classical} limit $T\gg\frac{1}{C}$ in which the quantum effects are much smaller than the temperature scale, and then the opposite 
\emph{quantum} regime  $T\ll\frac{1}{C}$ in which quantum effects dominate over temperature. 
However, one should keep in mind that the results obtained in the quantum regime should be interpreted with caution, as a new theoretical framework may be needed for a proper description of hydrodynamics and of the shear viscosity (we return to this point in the Conclusions). 
Nonetheless, we present our $\eta/s$ results for $T C \ll 1$ 
since they may still offer useful physical intuition, with the understanding that a more thorough analysis could lead to a different behavior in this particular temperature range.
Throughout the analysis we focus on the regime $\omega \ll M$, which implies\footnote{One can show that $M\propto T^2$ and $M\propto T$ at the saddle points, in the semiclassical and quantum regimes respectively.} $\omega \ll T$, ensuring that the frequency remains the lowest energy scale in the system and that hydrodynamics is well-defined (subject to the caveat above).

In the two temperature ranges, we have:
\begin{itemize}
\item Semi-classical regime, $T\gg\frac{1}{C}$ and $\omega\ll M$:\\
Note that we can approximate the momenta as follows,
\begin{align}
&k_1\simeq\sqrt{2CM}+\frac{\omega}{2}\sqrt{\frac{2C}{M}},\quad k_2=\sqrt{2CM} \, , \nonumber\\
&\Rightarrow\quad dk_1dk_2=\frac{1}{4M}(2C)\, dM d\omega\, .
\end{align}
By taking $k_1\geq0$ and $k_2\geq0$ and dropping terms proportional to $\omega/M$, the integral \eqref{eq:g1} can be approximated as
\begin{align}
\langle \mathcal{G}(t)\rangle & \approx\frac{e^{S_0}}{Z(\beta)} \frac{4}{\pi^2(2C)^{2\ell-2}}\int \frac{1}{16}dMd\omega e^{4\pi\sqrt{2CM}+\pi\omega\sqrt{\frac{2C}{M}}}e^{-i\omega t-M\beta}\nonumber\\
& \quad\times \frac{1}{\Gamma(2\ell)}\, \Gamma \left(\ell+i\left(2\sqrt{2CM}+\frac{\omega}{2}\sqrt{\frac{2C}{M}}\right)\right)\, \Gamma \left(\ell-i\left(2\sqrt{2CM}+\frac{\omega}{2}\sqrt{\frac{2C}{M}}\right)\right)\nonumber\\
& \quad\times \Gamma \left(\ell+i\left(\frac{\omega}{2}\sqrt{\frac{2C}{M}}\right)\right)\, \Gamma \left(\ell-i\left(\frac{\omega}{2}\sqrt{\frac{2C}{M}}\right)\right)\, .\label{eq:g2}
\end{align}

Since $\ell$ is of the same order as the scalar field  mass, \emph{i.e.} $\ell\ll M$, the first two Gamma functions can be approximated by
\begin{align}
 {}& \Gamma\left(\ell+i\left(2\sqrt{2CM}+\frac{\omega}{2}\sqrt{\frac{2C}{M}}\right)\right)\, \Gamma\left(\ell-i\left(2\sqrt{2CM}+\frac{\omega}{2}\sqrt{\frac{2C}{M}}\right)\right)\nonumber \\
  &\approx\frac{2\pi\left(2\sqrt{2CM}\right)^{2\ell-1}}{e^{\pi\left(2\sqrt{2CM}+\frac{\omega}{2}\sqrt{\frac{2C}{M}}\right)}}\, ,
\end{align}
where we used the identity
\begin{align}
  {}&\Gamma\left(\ell+i\left(2\sqrt{2CM}+\frac{\omega}{2}\sqrt{\frac{2C}{M}}\right)\right)\, \Gamma\left(\ell-i\left(2\sqrt{2CM}+\frac{\omega}{2}\sqrt{\frac{2C}{M}}\right)\right)\nonumber\\
 & =\frac{\pi\left(2\sqrt{2CM}+\frac{\omega}{2}\sqrt{\frac{2C}{M}}\right)}{{\rm sinh}\pi\left(2\sqrt{2CM}+\frac{\omega}{2}\sqrt{\frac{2C}{M}}\right)}\prod^{\ell-1}_{n=1}\left(n^2+\left(2\sqrt{2CM}+\frac{\omega}{2}\sqrt{\frac{2C}{M}}\right)^2\right)\, .
\end{align}
We evaluate this integral using the saddle-point approximation and a  Fourier transformation to obtain the quantum-corrected retarded Green's function

\begin{align}
\hspace*{-1cm}\langle\mathcal{G}_R(\omega,\beta)\rangle &=i\frac{1}{2}\left(1-e^{-\beta\omega}\right)\nonumber\\
&\times\frac{(2\pi)^{2\ell-1}}{{\rm sin}\left(\pi\ell + \pi i\frac{\omega}{2\pi}\times2C \left[\frac{\beta}{2C} + \frac{(1-2\ell)}{2\pi^2} \left(\frac{\beta}{2C}\right)^2\right] \right)}\textrm{exp}\left[\frac{\omega}{2}\times2C\left(\frac{\beta}{2C}+\frac{1-2\ell}{2\pi^2}\left(\frac{\beta}{2C}\right)^2\right)\right]\nonumber\\
 &\times\left[2C\left(\frac{\beta}{2C}+\frac{1-2\ell}{2\pi^2}\left(\frac{\beta}{2C}\right)^2\right)\right]^{1-2\ell}\frac{\Gamma \left(\ell-i\frac{\omega}{2\pi}\times2C \left[\frac{\beta}{2C} + \frac{(1-2\ell)}{2\pi^2} \left(\frac{\beta}{2C}\right)^2\right]\right)}{\Gamma \left(1-\ell-i\frac{\omega}{2\pi}\times2C \left[\frac{\beta}{2C} + \frac{(1-2\ell)}{2\pi^2} \left(\frac{\beta}{2C}\right)^2\right]\right)}\, .\label{eq:c1}
\end{align}

The derivation follows fairly standard steps, which we describe in detail in Appendix~\ref{Sec:AppendixA1}.
Comparing the quantum-corrected Green's function \eqref{eq:c1} to the tree-level Green's function \eqref{eq:ads2tree}, we interpret the terms that are higher orders in $\beta$ as encoding the quantum corrections coming from the gravity fluctuations.

\item Quantum regime, $T\ll\frac{1}{C}$ and $\omega\ll M$:\\

For the $CT\ll1$ case, we need to reconsider Eq.~\eqref{eq:g1}. Since $\ell$ is of the order of the particle mass and $CM$ is extremely small, we shall take the limit $\ell\gg CM$. Then Eq.~\eqref{eq:g1} becomes
\begin{align}
\langle \mathcal{G}(t)\rangle &\approx\frac{e^{S_0}}{Z(\beta)} \frac{1}{\pi^2(2C)^{2\ell-2}}\frac{(\ell-1)!^2}{\Gamma(2\ell)}4\pi^2\int d\omega e^{-i\omega t}\int dM\; 2CMe^{\pi\frac{\omega}{2}\sqrt{\frac{2C}{M}}}e^{-M\beta}\, \nonumber\\
&\times  \Gamma \left(\ell+i\left(\frac{\omega}{2}\sqrt{\frac{2C}{M}}\right)\right)\, \Gamma \left(\ell-i\left(\frac{\omega}{2}\sqrt{\frac{2C}{M}}\right)\right)\, ,\label{eq:b1}
\end{align}
where we have used
\begin{align}
{}&\Gamma\left(\ell+i\left(2\sqrt{2CM}+\frac{\omega}{2}\sqrt{\frac{2C}{M}}\right)\right)\, \Gamma\left(\ell-i\left(2\sqrt{2CM}+\frac{\omega}{2}\sqrt{\frac{2C}{M}}\right)\right)\,\nonumber \\
  &\approx \frac{\pi\left(2\sqrt{2CM}+\frac{\omega}{2}\sqrt{\frac{2C}{M}}\right)}{{\rm sinh}\pi\left(2\sqrt{2CM}+\frac{\omega}{2}\sqrt{\frac{2C}{M}}\right)}(\ell-1)!^2\, ,
\end{align}
and we have taken the limit $\omega\ll M\ll\frac{1}{C}$.

We evaluate this integral using the saddle-point approximation and performing a Fourier transformation, after which we obtain the
following quantum-corrected retarded Green's function
\begin{align}
\langle\mathcal{G}_R(\omega,\beta)\rangle &=i\frac{1}{2}\left(1-e^{-\beta\omega}\right)\frac{(\ell-1)!^2}{2^{-5/2}e\pi}\frac{(2C)^{1-2\ell}}{\Gamma(2\ell)}\left(\frac{\beta}{2C}\right)^{-\frac{1}{2}}  \nonumber\\
&\times e^{\frac{\omega}{2}\pi\sqrt{2C\beta}}\frac{\pi}{{\rm sin}(\pi\ell + \pi i\frac{\omega}{2}\sqrt{2C\beta} )} \frac{ \Gamma \left(\ell-i\frac{\omega}{2\pi}\pi\sqrt{2C\beta}\right)}{\Gamma(1-\ell-i\frac{\omega}{2}\sqrt{2C\beta})}\,.\label{eq:c2}
\end{align}
Details of the derivation are described in Appendix~\ref{Sec:AppendixA2}. 
Also, in Appendix~\ref{Sec:AppendixA3} we discuss a subtlety that arises from 
the fact that in this limit the 
Wightman Green's function displays a period that is not set by $\beta$, but rather $\pi\sqrt{2C\beta}$.
This in turn has implications for the KMS condition.

\end{itemize}

\subsubsection{A quantum corrected $\ell'$}

For a more compact way to encode the difference between the tree-level and quantum cases, it is useful to rewrite 
the quantum-corrected Green’s function $\langle \mathcal{G}_R(\omega,T)\rangle$ in the same form as the tree-level Green’s function $\mathcal{G}_R(\omega,T)$, which we repeat here for convenience,
\begin{equation}
\label{treelevelsection4}
\mathcal{G}_R(\omega,T)=-(\pi T)^{2\ell-1}\frac{\Gamma(\ell-\frac{i\omega}{2\pi T})}{\Gamma(1-\ell-\frac{i\omega}{2\pi T})}\frac{\Gamma(\frac{3}{2}-\ell)}{\Gamma(\frac{1}{2}+\ell)}\, .
\end{equation}
As we will see, this will entail introducing a new \emph{renormalized parameter}  $\ell'$ which encodes the 
non-trivial dependence on the quantum effects and is generically temperature dependent. Deep in the classical regime, $\ell'$ reduces to $\ell$ as expected. While rewriting the Green's function in this way is not necessary, it is convenient and makes the final results compact, as we will see.

As in the previous discussion, we examine the two temperature regimes separately:
\begin{itemize}
\item Semi-classical regime $T\gg\frac{1}{C}$:\\
The total quantum corrected Green's function is 
\begin{align}
\hspace*{-1cm}\langle\mathcal{G}_R(\omega,\beta)\rangle &=i\frac{1}{2}\left(1-e^{-\beta\omega}\right)\nonumber\\
&\times\frac{(2\pi)^{2\ell-1}}{{\rm sin}\left(\pi\ell + \pi i\frac{\omega}{2\pi}\times2C \left[\frac{\beta}{2C} + \frac{(1-2\ell)}{2\pi^2} \left(\frac{\beta}{2C}\right)^2\right] \right)}\textrm{exp}\left[\frac{\omega}{2}\times2C\left(\frac{\beta}{2C}+\frac{1-2\ell}{2\pi^2}\left(\frac{\beta}{2C}\right)^2\right)\right]\nonumber\\
 &\times\left[2C\left(\frac{\beta}{2C}+\frac{1-2\ell}{2\pi^2}\left(\frac{\beta}{2C}\right)^2\right)\right]^{1-2\ell}\frac{\Gamma \left(\ell-i\frac{\omega}{2\pi}\times2C \left[\frac{\beta}{2C} + \frac{(1-2\ell)}{2\pi^2} \left(\frac{\beta}{2C}\right)^2\right]\right)}{\Gamma \left(1-\ell-i\frac{\omega}{2\pi}\times2C \left[\frac{\beta}{2C} + \frac{(1-2\ell)}{2\pi^2} \left(\frac{\beta}{2C}\right)^2\right]\right)}\, .\label{eq:complete1}
\end{align}
Let's rewrite Eq.\,(\ref{treelevelsection4}) and Eq.\,(\ref{eq:complete1})  in the following form,
\begin{align}
\mathcal{G}_R(\omega,\beta) & = F_1(\beta,\ell,\omega)\times \left(\frac{\beta}{2C}\right)^{1-2\ell}\, ,\nonumber\\
\langle \mathcal{G}_R(\omega,\beta)\rangle & = F_2(\beta,\ell,\omega)\times \left[\frac{\beta}{2C}+\frac{1-2\ell}{2\pi^2}\left(\frac{\beta}{2C}\right)^2\right]^{1-2\ell}\, ,
\end{align}
where
\begin{align}
\label{F1beta}
\hspace*{-1cm} F_1(\beta,\ell,\omega) &= -\frac{\Gamma(\frac{3}{2}-\ell)}{\Gamma(\frac{1}{2}+\ell)} (\pi)^{2\ell-1}(2C)^{1-2\ell}\frac{\Gamma(\ell-\frac{i\omega}{2\pi T})}{\Gamma(1-\ell-\frac{i\omega}{2\pi T})}\, ,\\
\hspace*{-1cm} F_2(\beta,\ell,\omega) &=i\frac{1}{2}\left(1-e^{-\beta\omega}\right)\frac{(2\pi)^{2\ell-1}(2C)^{1-2\ell}}{{\rm sin}\left(\pi\ell + \pi i\frac{\omega}{2\pi}\times2C \left[\frac{\beta}{2C} + \frac{(1-2\ell)}{2\pi^2} \left(\frac{\beta}{2C}\right)^2\right] \right)}\nonumber\\
&\times\textrm{exp}\left[\frac{\omega}{2}\times2C\left(\frac{\beta}{2C}+\frac{1-2\ell}{2\pi^2}\left(\frac{\beta}{2C}\right)^2\right)\right]\frac{\Gamma \left(\ell-i\frac{\omega}{2\pi}\times2C \left[\frac{\beta}{2C} + \frac{(1-2\ell)}{2\pi^2} \left(\frac{\beta}{2C}\right)^2\right]\right)}{\Gamma \left(1-\ell-i\frac{\omega}{2\pi}\times2C \left[\frac{\beta}{2C} + \frac{(1-2\ell)}{2\pi^2} \left(\frac{\beta}{2C}\right)^2\right]\right)}  \, .
\end{align}

Next, note that the quantum-corrected Green's function can be written in the same form as the tree-level Green's function,
\begin{align}
\langle \mathcal{G}_R (\omega,\beta)\rangle & = F_2(\beta,\ell,\omega)\times \left[\frac{\beta}{2C} + \frac{1-2\ell}{2\pi^2} \left(\frac{\beta}{2C}\right)^2\right]^{1-2\ell}\,,\nonumber\\
 & = F_1(\beta,\ell,\omega)\times \left(\frac{\beta}{2C}\right)^{1-2\ell_1'}\, ,\label{eq:Green Fct with F1}
\end{align}
by introducing the renormalized parameter  $\ell_1'$, which encodes the effects of quantum corrections on $\ell$. The new parameter has the following explicit expression,
\begin{equation}
\ell_1'=\frac{1}{2}-\frac{{\rm log} \left(\frac{F_2(\beta,\ell,\omega)}{F_1(\beta,\ell,\omega)}\right) + (1-2\ell)\, {\rm log} \left[\frac{\beta}{2C}+\frac{1-2\ell}{2\pi^2} \left(\frac{\beta}{2C}\right)^2\right]}{2\, {\rm log}\, \frac{\beta}{2C}}\, ,
\end{equation}
and is clearly temperature dependent. We plot $\ell'_1$ in Figure \ref{fig:ell1}, where we see that it approaches $\ell$ (straight blue line) at very high temperatures, corresponding to the tree level case.

\begin{figure}[htb!]
\begin{center}
\includegraphics[width=10cm, angle=0]{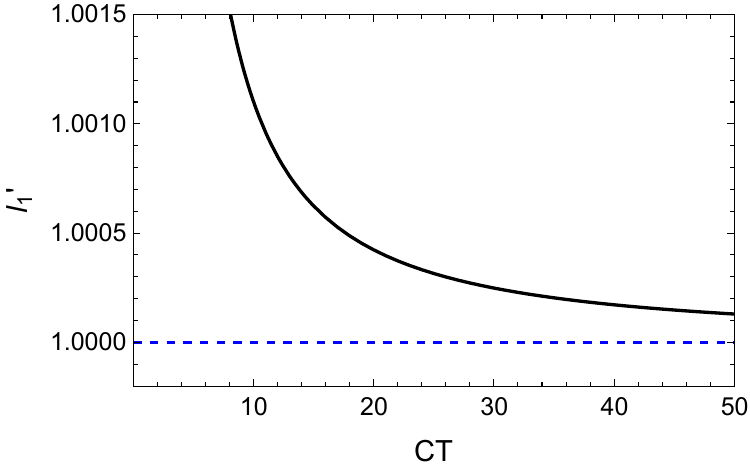}
\caption{The temperature dependence of the renormalized parameter $\ell'_1$ (solid curve) that encodes the quantum
corrections on the scaling dimension compared to $\ell=1$ (dashed line). We consider $T\gg1/C$ with $\kappa = 1$, $L = 0.1$, $r_0 = 1$.}\label{fig:ell1}
\end{center}
\end{figure}

\begin{figure}[htb!]
\begin{center}
\includegraphics[width=7.5cm, angle=0]{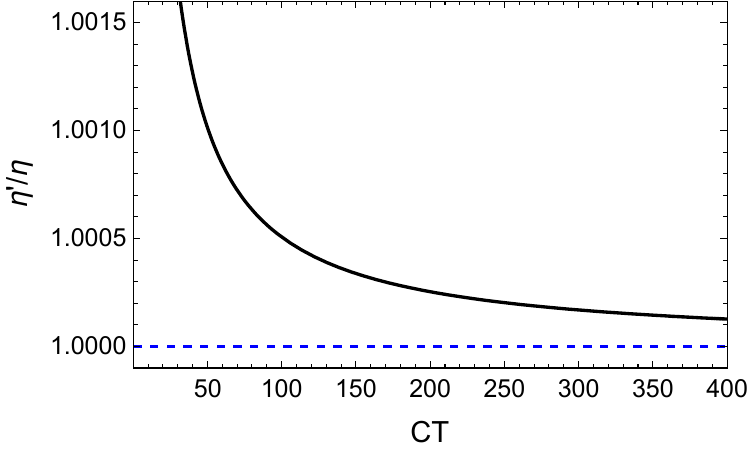}
\includegraphics[width=7.5cm, angle=0]{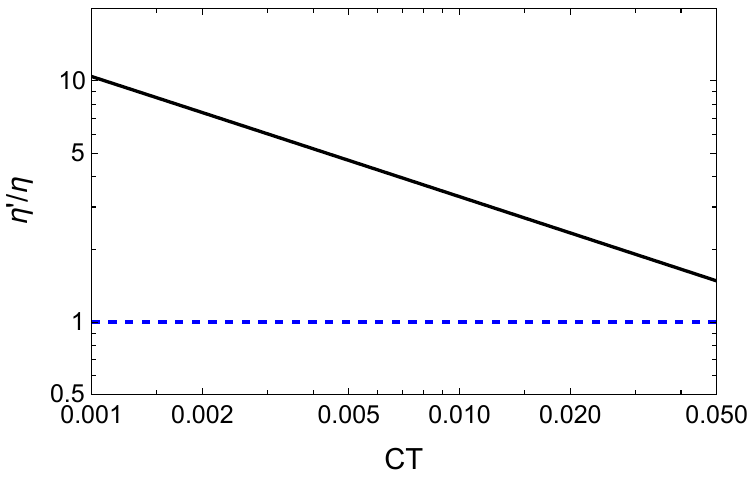}
\caption{Temperature dependence of $\eta'/\eta$ (solid curve) compared to its classical limit 
(dashed line). The left panel corresponds to $T\gg1/C$ while the right panel (log-log scale) to $T\ll1/C$. In both panels we have taken  $\kappa = 1$, $L = 0.1$, $r_0 = 1$.}\label{fig:etaprime}
\end{center}
\end{figure}

\begin{figure}[htb!]
\begin{center}
\includegraphics[width=10cm, angle=0]{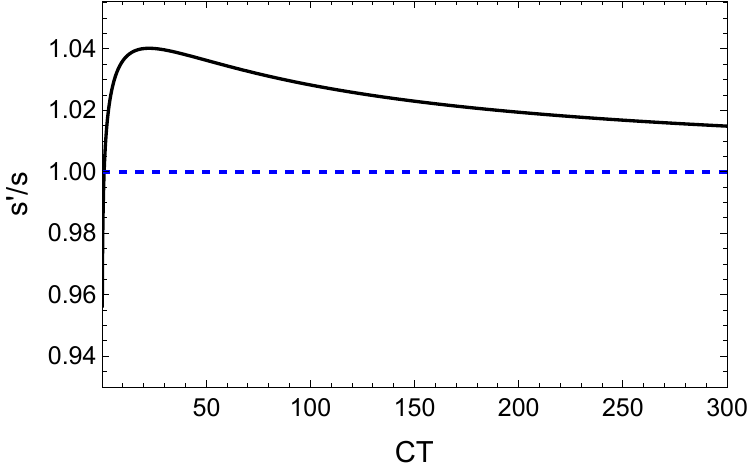}
\caption{Temperature dependence of $s'/s$ (solid curve) compared to its classical limit 
(dashed line) with $CT\gg1$ and $\kappa = 1$, $L = 0.1$, $r_0 = 1$.}\label{fig:sprime}
\end{center}
\end{figure}

\item Quantum regime $T\ll\frac{1}{C}$:\\
In this case the total quantum corrected Green's function is 
\begin{align}
\langle\mathcal{G}_R(\omega,\beta)\rangle &= i\frac{1}{2}\left(1-e^{-\beta\omega}\right)\frac{(\ell-1)!^2}{2^{-5/2}e\pi}\frac{(2C)^{1-2\ell}}{\Gamma(2\ell)}\left(\frac{\beta}{2C}\right)^{-\frac{1}{2}}  \nonumber\\
&\times e^{\frac{\omega}{2}\pi\sqrt{2C\beta}}\frac{\pi}{{\rm sin}(\pi\ell + \pi i\frac{\omega}{2}\sqrt{2C\beta} )} \frac{ \Gamma \left(\ell-i\frac{\omega}{2\pi}\pi\sqrt{2C\beta}\right)}{\Gamma(1-\ell-i\frac{\omega}{2}\sqrt{2C\beta})}\,.
\end{align}

Once again, the quantum-corrected Green’s function can be rewritten in the same form as the tree-level Green’s function,
\begin{align}
\mathcal{G}_R(\omega,\beta) & = F_1(\beta,\ell,\omega)\times \left(\frac{\beta}{2C}\right)^{1-2\ell}\, ,\nonumber\\
\label{GRF3}
\langle \mathcal{G}_R(\omega,\beta)\rangle & = F_3(\beta,\ell,\omega)\times \left(\frac{\beta}{2C}\right)^{-\frac{1}{2}}=F_1(\beta,\ell,\omega)\times \left(\frac{\beta}{2C}\right)^{1-2\ell_2'}\, ,
\end{align}
where $F_1(\beta,\ell,\omega)$ is defined as in Eq.\,\eqref{F1beta} and 
\begin{align}
\hspace*{-1cm}F_3(\beta,\ell,\omega) &=i\frac{1}{2}\left(1-e^{-\beta\omega}\right)\frac{(\ell-1)!^2}{2^{-5/2}e\pi}\frac{(2C)^{1-2\ell}}{\Gamma(2\ell)}  e^{\frac{\omega}{2}\pi\sqrt{2C\beta}}\frac{\pi}{{\rm sin}(\pi\ell + \pi i\frac{\omega}{2}\sqrt{2C\beta} )} \frac{ \Gamma \left(\ell-i\frac{\omega}{2\pi}\pi\sqrt{2C\beta}\right)}{\Gamma(1-\ell-i\frac{\omega}{2}\sqrt{2C\beta})} \, .
\end{align}

The renormalized parameter $\ell'$ introduced in Eq.\,\eqref{GRF3} is now given by
\begin{equation}
\ell_2' = \frac{3}{4}+\frac{1}{2}\frac{\log{\left(\frac{F_1(\beta,\ell,\omega)}{F_3(\beta,\ell,\omega)}\right)}}{\log{\left(\frac{\beta}{2C}\right)}} \, .
\end{equation}
\end{itemize}
We will make use of these expressions next, to compute the shear viscosity.

\subsection{Quantum corrected shear viscosity and entropy density}
\label{Section43}

By considering quantum fluctuations, the relation~\eqref{eq:2d4d} between the
UV and IR Green's functions becomes
\begin{equation}
\langle G_R(\omega,T)\rangle=-\frac{1}{2\kappa^2}\langle\left(\frac{r_h}{L}\right)^2\rangle \langle\mathcal{G}_R(\omega,T)\rangle \, .
\end{equation}
As discussed in Section \ref{Section41}, since the near horizon geometry is AdS$_2\times\mathbb{R}^2$, the quantum fluctuation of $\langle\left(\frac{r_h}{L}\right)^2\rangle$ is controlled by the scale $M_{U(1)\times U(1)}$, while the quantum fluctuation of $\langle\mathcal{G}_R(\omega,T)\rangle$ is controlled by $1/C$. In this manuscript, we focus on the regime~\eqref{eq:mc}, which means we can neglect the quantum fluctuation of $\langle\left(\frac{r_h}{L}\right)^2\rangle$ and treat it as a background. Thus, we take the quantum-corrected AdS$_2$/CFT$_1$ Green's function $\langle\mathcal{G}_R(\omega,T)\rangle$ and the quantum-corrected AdS$_4$/CFT$_3$ Green's function $\langle G_R(\omega,T)\rangle$ to be related by 
\begin{equation}
\langle G_R(\omega,T)\rangle=-\frac{1}{2\kappa^2}\left(\frac{r_h}{L}\right)^2 \langle\mathcal{G}_R(\omega,T)\rangle \, .
\end{equation}
Using Kubo's formula, 
the quantum corrected shear viscosity
$\eta'$ is then\footnote{For $\ell=1$, $F_1(\beta,\ell,\omega)=\frac{1}{2C}\frac{i\omega}{T}$.}
\begin{align}\label{eq:correcteta}
\eta' &= -\lim_{\omega\to0}\frac{1}{\omega}{\rm Im}[\langle G_R(\omega,T)\rangle]=  \lim_{\omega\to0}\frac{1}{\omega}{\rm Im}\left[\frac{1}{2\kappa^2}\left(\frac{r_h}{L}\right)^2\left(\frac{i\omega}{T}\times T^{2\ell'-1}\right)\right]\,,  \nonumber\\
&=\frac{1}{2\kappa^2}\left(\frac{r_h}{L}\right)^2T^{2\ell'-2} \ ,
\end{align}
where the temperature dependent $\ell'$ exponent is $\ell'=\ell'_1$ for $T\gg\frac{1}{C}$ and $\ell'=\ell'_2$ for $T\ll\frac{1}{C}$  
(also recall that $r_h$ itself carries temperature dependence).
It is now clear why introducing the parameter $\ell'$ is useful -- in general, the quantum-corrected expression for the viscosity is quite involved, but $\ell'$ allows us to write it in a compact  way.

To better illustrate the behavior of the shear viscosity, we write down its approximated expression at high temperature $T\gg\frac{1}{C}$ for $\ell =1$, 
\begin{align}\label{eq:appa}
\eta'=\frac{1}{2\kappa^2}\left(\frac{r_h}{L}\right)^2T^{2\ell_1'-2}&\approx\frac{1}{2\kappa^2}(\frac{r_h}{L})^2\left(1+\frac{1}{4 \pi ^2}(CT)^{-1}+\cdots\right)\\
&\approx\frac{1}{2\kappa^2}\left(\frac{r_0}{L}\right)^2\left(1+\frac{2 \pi  L^2}{3 r_0}T+\frac{1}{4 \pi ^2}(CT)^{-1}+\cdots\right)\, ,\label{eq:app1}
\end{align}
where $\cdots$ denotes $\mathcal{O}(r_0^{-2})$ and $\mathcal{O}((CT)^{-2})$ terms.
We note that this expression \eqref{eq:app1} for the quantum-corrected $\eta'$ in the semiclassical region is consistent with the complementary work \cite{PandoZayas:2025snm,Kanargias:2025vul}. However, the full expression \eqref{eq:correcteta} is generically more involved, and captures additional effects to those that appear to leading order in the expansion \eqref{eq:app1}.

On the other hand, in the quantum regime $T\ll\frac{1}{C}$, taking
 $\ell=1$  we have
\begin{align}\label{eq:appb}
\eta'=\frac{1}{2\kappa^2}\left(\frac{r_h}{L}\right)^2T^{2\ell_2'-2}&=\frac{1}{2\kappa^2}\left(\frac{r_h}{L}\right)^2\times \frac{2^{3/2}}{e\pi}\frac{1}{\sqrt{2CT}}\\
&\approx \frac{1}{2\kappa^2}\left(\frac{r_0}{L}\right)^2\times \frac{2^{3/2}}{e\pi}\frac{1}{\sqrt{2CT}}+\cdots \, .\label{eq:app2}
\end{align}
We will come back to these expressions for the shear viscosity in the next section, where we make a direct comparison to the behavior of the absorption cross-section.
Also, we should stress that when we plot the viscosity to entropy ratio we will use 
the full expression \eqref{eq:correcteta} for the shear viscosity, and not the approximations above. 
In Figure \ref{fig:etaprime}
we plot the ratio of the quantum-corrected shear viscosity $\eta'$ to the classical shear viscosity $\eta$ in the two distinct temperature regimes.

Since we are ultimately interested in $\eta/s$, we will need the expression for the entropy.
Following \cite{Mertens:2022irh,Iliesiu:2020qvm}, by considering the one-loop quantum correction from JT gravity, the quantum-corrected entropy density $s'$ is given by
\begin{equation}
\label{sprime}
s' = \frac{2\pi}{\kappa^2}\left(\frac{r_h}{L}\right)^2 + \frac{3}{8\pi L^2}\log(CT) \, .
\end{equation}
One should keep in mind that, because of the log term, this expression breaks down at some (small) value of $CT$, where the entropy becomes negative \cite{Turiaci:2023wrh}. 
Thus, when using this expression, we shouldn't take $CT$ arbitrarily small.
However, it has been shown that the 
negative entropy problem can be remedied by the inclusion of wormholes in a nonperturbative completion of JT gravity \cite{Johnson:2019eik,Johnson:2020exp,Johnson:2020mwi,Johnson:2021rsh,Johnson:2021zuo,Johnson:2021tnl,Johnson:2022wsr,Johnson:2022pou}, and 
that the behavior of the entropy can be traced to whether the free energy is quenched or annealed (see \emph{e.g.} the discussion in \cite{Engelhardt:2020qpv,Chandrasekaran:2022asa,Hernandez-Cuenca:2024icn}).
In this paper we will always restrict our analysis to the regime in which  \eqref{sprime} is positive, and ignore these contributions. 

We plot the ratio of the quantum-corrected entropy density $s'$ to the classical entropy density $s$ in Figure \ref{fig:sprime}, focusing on the semi-classical temperature regime $CT\gg1$. Notice that the resulting curve is not monotonic, and indeed $s'/s$ reaches a maximum (around $TC \sim 22$). We will come back to this point shortly.

Using \eqref{sprime}, 
we finally obtain the quantum-corrected $\eta/s$, 
\begin{equation}
\frac{\eta'}{s'}=\frac{\frac{1}{2\kappa^2}(\frac{r_h}{L})^2T^{2\ell'-2}}{s + \frac{3}{8\pi L^2}\log(CT)} 
= \frac{1}{4\pi} \frac{T^{2\ell'-2}}{1+\frac{3\kappa^2}{16\pi^2r^2_0}\left(1+\frac{2 \pi  L^2}{3 r_0}T\right)^{-1}\log(CT)} \, ,
\end{equation}
where we emphasize that additional temperature dependence is hidden in $\ell'$.

\begin{figure}[htb!]
\begin{center}
\includegraphics[width=12cm, angle=0]{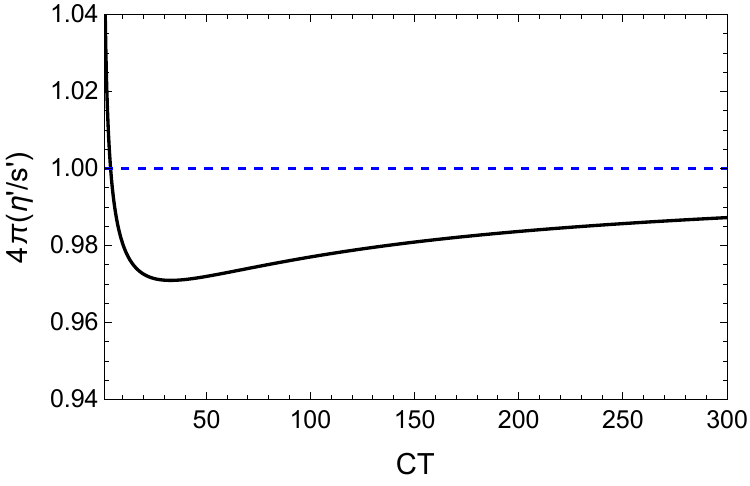}
\caption{
The quantum-corrected shear viscosity-entropy ratio as a function of temperature in the semiclassical regime $CT\gg 1$. The dashed blue line denotes the KSS bound. We have chosen $\kappa = 1$, $L = 0.1$, $r_0 = 1$.}\label{fig:ct1}
\end{center}
\end{figure}

\begin{figure}[htb!]
\begin{center}
\includegraphics[width=7.3cm, angle=0]{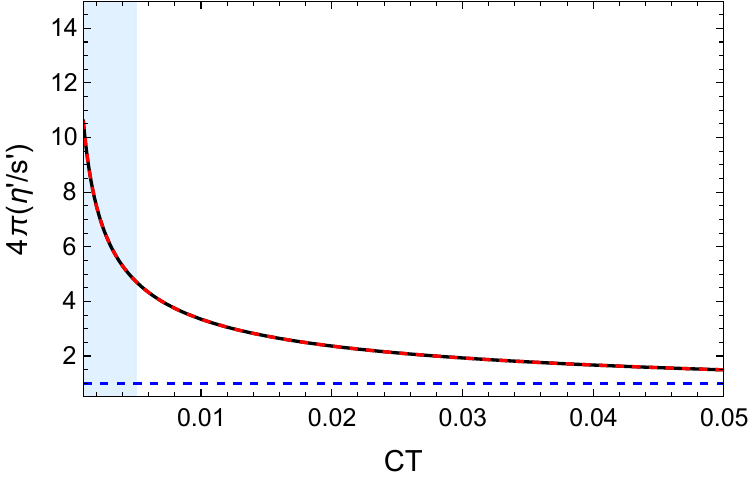}
\includegraphics[width=7.5cm, angle=0]{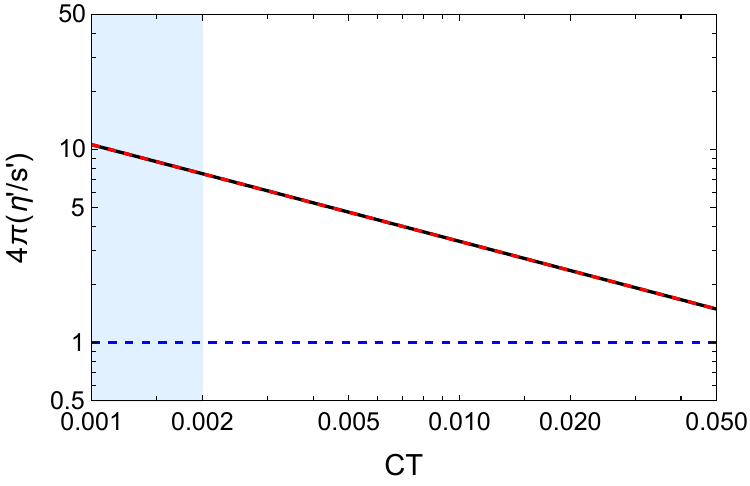}
\caption{The left panel shows the temperature dependence of the quantum-corrected shear viscosity-entropy ratio in the quantum regime $CT\ll 1$. The dashed blue line denotes the KSS bound and the dashed red line denoted the fitting function $4\pi(\eta'/s')\approx0.24(CT)^{-0.5}$. The right panel shows the same quantities in a log-log plot. The blue shaded area schematically represents the region where the entropy is negative, indicating the breakdown of the quantum-corrected $\eta/s$. We have set $\kappa = 1$, $L = 0.1$, $r_0 = 1$.}\label{fig:ct2}
\end{center}
\end{figure}

Inspecting the semi-classical limit and quantum limits, we have
\begin{itemize}
\item Semi-classical regime $T\gg\frac{1}{C}$:
\begin{equation}
\label{finaletaslargeT}
\boxed{\frac{\eta'}{s'}= \frac{1}{4\pi} \frac{T^{2\ell_1'-2}}{1+\frac{3\kappa^2}{16\pi^2r^2_0}\left(1+\frac{2 \pi  L^2}{3 r_0}T\right)^{-1}\log(CT)}\,. }
\end{equation}
The behavior as a function of temperature is shown in Figure \ref{fig:ct1}, where the horizontal (blue) line denotes the universal result. First, we note that for very high $CT$, one recovers $1/4\pi$ as expected. Then, as 
the temperature is lowered, $\eta/s$ first decreases and violates the KSS bound, then reaches a minimum value, and subsequently increases, eventually exceeding the KSS bound and rising significantly above it. 

It is interesting that $\eta/s$ attains a minumum in the semi-classical region (and not, as we are about to see, deep in the quantum region). 
It is easy to see from Figure \ref{fig:sprime} that this  comes from the maximum of the entropy, and not from the shear viscosity itself, which is monotonic (see Figure \ref{fig:etaprime}, left panel).
The critical temperature for the inflection point (the minimum $\eta/s$) is given by
\begin{equation}
\label{critT}
\partial_T\left(\frac{\eta'}{s'}\right)=0\rightarrow T_c\approx \frac{3 r_0}{2 \pi  L^2}\frac{1}{W\left(\frac{3 C r_0}{2 e \pi  L^2}\right)}\, ,
\end{equation}
where $W$ is the Lambert-W function and $e$ is Euler's number.
All we can say at this stage is that the minimum is indeed generated by the presence of quantum corrections 
to the entropy in the semi-classical regime.

Finally, since the full expression \eqref{finaletaslargeT} hides some of the temperature dependence inside the parameter $\ell_1'$, it may be useful to further approximate it at very large temperatures. Doing so yields the 
following expression,
\begin{equation}
\frac{\eta'}{s'}= \frac{1}{4\pi} \frac{1+\frac{1}{4 \pi ^2}(CT)^{-1}+\frac{1 }{16 \pi ^4}(CT)^{-2}+\cdots}{1+\frac{3\kappa^2}{16\pi^2r^2_0}\left(1+\frac{2 \pi  L^2}{3 r_0}T\right)^{-1}\log(CT)}\,,
\end{equation}
with all temperature dependence now explicit. 
This approximation can be compared to Eq.\,(4.30) of \cite{PandoZayas:2025snm}, 
up to a Taylor expansion in temperature and $r_0$.

\item Quantum regime $T\ll\frac{1}{C}$:
Similarly, in the opposite regime we have 
\begin{equation}
\frac{\eta'}{s'} = \frac{1}{4\pi} \frac{T^{2\ell_2'-2}}{1+\frac{3\kappa^2}{16\pi^2r^2_0}\left(1+\frac{2 \pi  L^2}{3 r_0}T\right)^{-1}\log(CT)} \, .
\end{equation}
The temperature dependence is shown in Figure \ref{fig:ct2}, where the straight (blue) line once again represents $1/4\pi$. As the temperature is lowered, $\eta/s$ increases rapidly and its value becomes much larger than that of the semi-classical regime shown in Figure~\ref{fig:ct1}. This behavior is consistent with our prediction that $\eta/s$ should continue to increase in the quantum region. However, it is important to note that (because of the breakdown of the entropy) we can not predict what will happen at temperatures close to zero.
\end{itemize}

\section{Comparison to the absorption cross-section}
\label{Section5}

Recall that the absorption cross section $\sigma$ of a graviton incident on a black brane with energy $\omega$, and polarized parallel to the brane (e.g. along the xy directions), is related to the field theory stress-tensor correlator via \cite{Klebanov:1997kc,Gubser:1997yh,Policastro:2001yc}
\begin{equation}
\label{abscross}
\sigma(\omega) = \frac{\kappa^2}{\omega}\int dtd\vec{x}e^{i\omega t}\langle|T_{xy}(t,\vec{x}),T_{xy}(0,0)|\rangle\,.
\end{equation}
Comparison with \eqref{etastress} 
shows that in the low-frequency limit the cross section can be related directly to the shear viscosity,
\begin{equation}
\label{etasigmarelation}
    \eta = \frac{1}{2\kappa^2}\sigma (\omega =0)\,.
\end{equation}
Now that we have an expression for the quantum-corrected shear viscosity, we can examine how it compares to the absorption cross-section, and whether the simple relation \eqref{etasigmarelation} still holds.
Following \cite{Emparan:2025sao} (see also \cite{Biggs:2025nzs}), in order to compute the  absorption cross-section obtained by scattering a low-frequency wave of a massless, minimally coupled scalar off a 4D black hole, 
one can plug the absorption and emission rates per unit frequency \cite{Mertens:2019bvy,Blommaert:2020yeo} into the relation for the absorption cross-section with the rates \cite{Das:1996we,Gubser:1996zp} and get
\begin{equation}\label{eq:sigma1}
\sigma_{\text{abs}} (\omega) =
\frac{4\pi r_0^2}{\omega}
\left(|\langle E_i + \omega | \mathcal{O} | E_i \rangle|^2 \rho(E_i + \omega)-|\langle E_i - \omega | \mathcal{O} | E_i \rangle|^2 \rho(E_i - \omega)
\right) \, , 
\end{equation}
where $E_i$ is the initial energy of black hole, $\mathcal{O}$ corresponds to the operator dual to the massless scalar field and $\rho$ is the density of black hole states. 
The papers \cite{Mertens:2022irh,Stanford:2017thb,Mertens:2017mtv,Kitaev:2018wpr,Yang:2018gdb,Iliesiu:2019xuh}  calculated the density of black hole  states and the matrix elements 
of conformal primaries 
in comprehensive studies of the Schwarzian theory and its coupling to conformal matter. Substituting these results into Eq.\,\eqref{eq:sigma1},  the  absorption cross-section\footnote{It should be straightforward to generalize the black hole results of \cite{Emparan:2025sao} to a black brane.} in the microcanonical ensemble reads~\cite{Emparan:2025sao}
\begin{equation}\label{eq:sigma2}
\sigma^{\rm mic}_{\rm abs}(\omega) = A_H \left(
\frac{\sinh\left( 2\pi \sqrt{\frac{2(E_i+\omega)}{E_b}} \right)}{
\cosh\left( 2\pi \sqrt{\frac{2(E_i+\omega)}{E_b}} \right)-\cosh\left( 2\pi\sqrt{\frac{2E_i}{E_b}} \right)}-\frac{
\sinh\left( 2\pi \sqrt{\frac{2(E_i-\omega)}{E_b}} \right)\Theta(E_i-\omega)}{\cosh\left( 2\pi \sqrt{\frac{2E_i}{E_b}} \right)-\cosh\left( 2\pi \sqrt{\frac{2(E_i-\omega)}{E_b}} \right)}\right) \, ,
\end{equation}
where $A_H=4\pi r^2_0$ and the quantum scale is $E_b=\frac{1}{C}$.
At very low frequencies with  the ratio $E_i/E_b$ fixed, Eq.\,\eqref{eq:sigma2} reduces to
\begin{equation}\label{eq:sigma3}
\sigma_{\text{abs}}^{\rm mic}(\omega \rightarrow 0) \rightarrow A_H \left( \coth \left( 2\pi \sqrt{\frac{2E_i}{E_b}} \right) + \frac{1}{2\pi} \sqrt{\frac{E_b}{2E_i}} \right) \, . 
\end{equation}
We need to consider the  following regimes:
\begin{itemize}
\item the semiclassical limit $E_i\ll E_b$, for which Eq.\,\eqref{eq:sigma3} becomes
\begin{equation}
\sigma_{\text{abs}}^{\rm mic}(\omega \rightarrow 0) \simeq A_H \left( 1 + \frac{1}{2\pi} \sqrt{\frac{E_b}{2E_i}} + 2 \exp \left( -4\pi \sqrt{\frac{2E_i}{E_b}} \right) + \cdots \right). 
\end{equation}
\item the quantum limit $E_i\gg E_b$, for which Eq.\,\eqref{eq:sigma3} becomes
\begin{equation}
\sigma_{\text{abs}}^{\rm mic}(\omega \rightarrow 0) \simeq A_H \frac{1}{\pi} \sqrt{\frac{E_b}{2E_i}}. 
\end{equation}
\end{itemize}
To compare our shear viscosity results to the absorption cross-section, we should convert the microcanonical absorption cross-section of \cite{Emparan:2025sao} to the canonical one. 
This can be done as follows: 
\begin{itemize}
\item In the semiclassical limit $E_i\gg E_b$  ($CT \gg 1$):
\begin{align}
\sigma_{\text{abs}}^{\rm can}(\omega) &= \frac{1}{Z(\beta)} \int_{0}^{\infty} dE \, \rho(E) e^{-\beta E} \sigma_{\text{abs}}^{\rm mic}\nonumber\\
&\approx A_H\left( 1+\frac{1}{4\pi^2CT}+\cdots\right) ,
\label{sigmacanclass}
\end{align}
where the black hole density of states $\rho(E)$ and the partition function $Z(\beta)$ are given by 
\begin{align}
\rho(E) &= \frac{1}{2\pi^2 E_{b}}e^{S_0}  \sinh\left( 2\pi \sqrt{\frac{2E}{E_{b}}} \right) \approx \frac{e^{S_0}}{4\pi^2 E_{b}}  \exp\left( 2\pi \sqrt{\frac{2E}{E_{b}}} \right), \\
Z(\beta) &= \int_{0}^{\infty} dE \, \rho(E) e^{-\beta E}
\approx \frac{e^{S_0}}{E_b^{3/2} \sqrt{2 \pi } \beta^{3/2}} \exp \left(\frac{2 \pi ^2}{E_b \beta}\right),
\end{align}
and we have used the relation $E_b=C^{-1}$.
We note that (\ref{sigmacanclass}) is consistent with the results of \cite{Biggs:2025nzs}.

To leading order in $1/CT$, 
this behavior is consistent with the temperature dependence obtained from the approximate expression~\eqref{eq:appa} of $\eta'$ in the semiclassical region $T \gg \frac{1}{C}$.
\item In the quantum limit $E_i\ll E_b$ ($CT \ll 1$):
\begin{align}
\sigma_{\text{abs}}^{\rm can}(\omega) &= \frac{1}{Z(\beta)} \int_{0}^{\infty} dE \, \rho(E) e^{-\beta E} \sigma_{\text{abs}}^{\rm mic}\nonumber\\
&\approx A_H\frac{1}{\pi}\sqrt{\frac{2}{\pi CT}} \,,
\label{crosscanqm}
\end{align}
where
\begin{equation}
\rho(E)\approx \frac{1}{\pi E_{b}} \left(  \sqrt{\frac{2E}{E_{b}}} \right)\,,\quad Z(\beta) \approx  \frac{1}{\sqrt{2\pi}}E_{brk}^{-3/2}\beta^{-3/2} \,.
\end{equation}Once again, (\ref{crosscanqm}) matches the results of \cite{Biggs:2025nzs}.
The expression for the cross section is consistent with the temperature dependence obtained from the approximate expression~\eqref{eq:appb} for the shear viscosity $\eta'$ in the quantum region $T \ll \frac{1}{C}$.
\end{itemize}
Even though \cite{Emparan:2025sao} worked with a near extremal black hole, the overall 
area $A_H$ used there is expressed in terms of the extremal horizon, for simplicity. However, one can easily generalize the result  to the near-extremal case. 

\vspace{-1mm}

\section{Conclusions}

In this paper our goal was to examine how the quantum corrections that modify the thermodynamics of black holes with an AdS$_2$ near-horizon geometry
affect the low-temperature behavior of $\eta/s$. 
Indeed, such corrections are generically expected to impact dynamics as well and -- in holographic models -- the transport properties of the dual system. 
In our analysis we have considered two distinct regimes, characterized by the competition between the temperature scale $T$ and the quantum scale $\sim \frac{1}{C}$ arising from the near-extremeal IR geometry. 
One is the semi-classical regime $CT \gg 1$, in which the temperature dominates over quantum effects, and the other is the quantum regime  $CT \ll 1$, in which the opposite is true.

Perhaps the most interesting feature we have observed -- a minimum for $\eta/s$ -- arises in the semiclassical regime, where the effects of strong quantum fluctuations can be safely neglected.
In this temperature range 
we find that $\eta/s$ deviates from the KSS bound, 
becoming smaller as $T$ is lowered, until it reaches a minimum at a critical temperature $T_c$ given in \eqref{critT}.
The shear viscosity itself in this region is monotonic. The minimum of $\eta/s$ can be explained by examining the ratio of the quantum corrected entropy $s'$ to the tree level entropy $s$. Indeed, $s'/s$ is not monotonic, and reaches a maximum in the semi-classical regime, precisely near $T_c$.
Below the critical temperature, $\eta/s$ starts to grow, and continues to do so rather quickly in the quantum phase. 
We also find that the temperature behavior of $\eta$ agrees with that of the black brane absorption cross section $\sigma$ computed in \cite{Emparan:2025sao,Biggs:2025nzs}, as expected 
on general grounds given the relation \eqref{etasigmarelation}.

One should keep in mind, however, that at extremely low temperatures the analysis is subtle and several issues and open questions arise:
\begin{itemize}
\item
First of all, 
as we have already mentioned, 
standard notions of hydrodynamics are likely to break down in the 
deep quantum regime (see e.g. \cite{Arean:2020eus}) at temperatures close to zero (see \cite{Blake:2017ris,Crossley:2015evo,Crossley:2015tka} for an approach to quantum hydrodynamics). 
However, \cite{Davison:2013bxa} found that a diffusion mode still exists even at zero temperature, and that hydrodynamics might be applicable (see also \cite{Edalati:2010pn}). 
In any case, our computations of $\eta$ in the quantum regime $CT\ll 1$ 
should be revisited, once we have a proper description of hydrodynamics there. It would be valuable
to understand to what extent our $\eta/s$ captures any of the key physics in the quantum phase.
\item
Also, as 
stressed in \cite{Emparan:2025sao}, when working within the naive hydrodynamic regime while requiring significant quantum fluctuations, the geometry itself may become sub-Planckian, which we don't know how to describe.  
Despite these challenges, which we did not attempt to address here, 
we have included the $CT \ll 1$ computation because it may help guide future holographic studies of transport. 

\item
Another issue is the fact that, 
at sufficiently low $CT$,
the quantum corrections to the black hole thermodynamic entropy cause it to become negative (see Section \ref{Section4} for a brief discussion of ways to obtain a positive entropy, and of the role of the quenched free energy). 
What this means is that we should not trust the expression \eqref{sprime} for the entropy, and thus $\eta/s$, in this particular temperature regime. 
Therefore -- until the behavior of the entropy is better understood -- one is unable to make any statement about $\eta/s$ very close to $T=0$.  
In particular, we can't compare the $CT \rightarrow 0$ limit of our computations to the result of \cite{Edalati:2009bi}, which obtained $\eta/s=1/4\pi$ exactly at $T=0$ using extremal charged black branes. 
This also ties together with the issues 
discussed above, regarding the 
quantum region $CT\ll1$. 

\item 
Another interesting point arises when we consider the KMS condition in the deep quantum regime.
The Wightman Green's function we obtained in \eqref{eq:b6} exhibits a modified period, indicating an emergent effective inverse temperature $\beta_{\rm eff} = \pi\sqrt{2C\beta}$. 
Since the periodicity of the Green's function usually determines the structure of the 
KMS condition, one is naively led to reformulate \eqref{eq:kms} in terms of $\beta_{\rm eff}$. However, doing so leads to an expression for the shear viscosity that does \emph{not} agree with the cross section in the deep quantum regime (see Appendix~\ref{Sec:AppendixA3} for details). It would be interesting to better understand the physical interpretation of this emergent effective temperature. 

\item
A point we should stress is that the minimum of $\eta/s$ is in the semi-classical region -- where the issues mentioned above are not relevant -- and \emph{not} in the deep quantum region. The question of what sets a fundamental lower bound for $\eta/s$ has been a longstanding one and is still open. In our analysis it is clear that the presence of the bulk quantum fluctuations -- representing finite $N$ effects in the IR CFT -- generates the temperature dependent flow that 
violates (in some regions of the temperature range) the KSS bound and leads to a minimum (with the entropy playing a crucial role in generating this minimum). 
Thus, it is tempting to argue that a fundamental lower bound to $\eta/s$
-- at least in holographic models -- must be tied to the interplay between finite $T$ effects and finite $N$ effects in the field theory. 
However, to truly address this question -- at least in this class of models -- one needs a better understanding of the low-$T$ regime of the theory and in particular of the entropy, as explained above.
We also note that the violation of the KSS bound seen here is partially reminiscent to the 
bound violations seen in holographic models with higher derivative corrections which involved $1/N$ corrections of the UV CFT \cite{Buchel:2008vz,Myers:2009ij,Cremonini:2009sy}. In this respect, the violation seen here is not unusual, but the mechanism that generates the temperature dependent flow is different.

\item
A particularly interesting and challenging open question  is 
the fate of $\eta$, 
and in turn $\eta/s$, exactly at zero temperature, once the effects of quantum fluctuations in the bulk are taken into account
within a proper hydrodynamic framework.
While we have computed the shear viscosity with quantum corrections in  both semiclassical $CT\gg1$ and quantum $CT\ll1$ regimes, it is also interesting to understand the transition between them. This can be done, in principle, by computing Eq.~\eqref{eq:g1} numerically.
Our analysis could also be extended to 
other transport coefficients of interest. 
Finally, we wonder if the absorption cross section 
can be recovered directly from the geometry -- appropriately modifying the metric to take into account quantum effects.
\end{itemize}
We leave these questions to future work.

\vspace{-1mm}

\section*{Acknowledgments}

We would like to thank Roberto Emparan, Leo Pando Zayas, Alex Buchel, Gustavo J. Turiaci, Hong Liu, Elias Kiritsis, Sameer Murthy, Blaise Goutéraux, Cong-Yuan Yue and Mohammad Moezzi for many helpful comments and discussions.
The work of S.C. was supported in part by the National Science
Foundation under Grant No.\,PHY-2210271.
L.L. was supported in part by the National Natural Science Foundation of China Grants No.\,12525503 and No.\,12447101.

\newpage
\appendix
\section{Quantum averaged Green's function from gravity and gauge fluctuations}
To evaluate the quantum averaged Green's function $\langle G_f (\omega) \rangle$ from gravity fluctuations, \eqref{eq:g2} and \eqref{eq:b1}, we use a saddle-point approximation and Fourier transformation. In this appendix, we present some details of the computation.

\subsection{Quantum Green's function for $T\gg\frac{1}{C}$}\label{Sec:AppendixA1}

We evaluate this integral using a saddle-point approximation. First, we rewrite the $M$-dependent terms in Eq.~\eqref{eq:g2} as a Gaussian integral:
\be\label{eq:Gaussian}
  \int dM\, e^{2\pi\sqrt{2CM} - M\beta + \pi\frac{\omega}{2}\sqrt{\frac{2C}{M}} + {\rm ln}\, \Gamma\left(\ell + i\frac{\omega}{2}\sqrt{\frac{2C}{M}}\right)+{\rm ln}\, \Gamma\left(\ell-i\frac{\omega}{2}\sqrt{\frac{2C}{M}}\right)+(\ell-\frac{1}{2})\, {\rm ln}\, 8CM} = \int dM\, e^{f(M)}\, ,
\ee
where
\begin{align}
f(M) = &2\pi\sqrt{2CM}-M\beta+\pi\frac{\omega}{2}\sqrt{\frac{2C}{M}}+{\rm ln}\Gamma\left(\ell+i\frac{\omega}{2}\sqrt{\frac{2C}{M}}\right) \nonumber\\
&+ {\rm ln}\Gamma\left(\ell-i\frac{\omega}{2}\sqrt{\frac{2C}{M}}\right) + \left(\ell-\frac{1}{2}\right){\rm ln}(8CM)\, .\label{eq:a2}
\end{align}
The saddle points are the solutions to the following equation:
\begin{align}
0 = \frac{\partial f(M)}{\partial M} & = \pi\sqrt{\frac{2C}{M}}-\beta-\frac{\pi}{4}\omega\sqrt{2C}M^{-3/2}-i\frac{\sqrt{2C}\omega}{4}M^{-3/2}\psi \left(\ell + i\frac{\sqrt C\omega}{\sqrt{M}} \right)\nonumber\\
& \quad + i\frac{\sqrt{2C}\omega}{4}M^{-3/2}\psi\left(\ell-i\frac{C\omega}{\sqrt{M}}\right)+\frac{\ell-\frac{1}{2}}{M}\, ,
\end{align}

where $\psi(x)$ is the digamma function, $\psi(z)=\frac{d}{dz}\ln \Gamma(z)$. In the limit $\omega\ll M$, the leading-order saddle-point solutions are given by
\begin{equation}\label{eq:sqrtm}
\frac{1}{\sqrt{2CM}}=\frac{-\pi\pm\sqrt{\pi^2+2(2\ell-1)\frac{\beta}{2C}}}{2\ell-1}\, .
\end{equation}
We expand the solution for positive $\sqrt{C M}$ and small $\beta$,
\begin{align}\label{eq:solution1}
\frac{1}{\sqrt{2CM}} & =\frac{-\pi+\sqrt{\pi^2+2(2\ell-1)\frac{\beta}{2C}}}{2\ell-1} = \frac{1}{\pi}\frac{\beta}{2C} + \frac{(1-2\ell)}{2\pi^3} \left(\frac{\beta}{2C}\right)^2 + O(\beta^3)\, .
\end{align}
Plugging it back into Eq.~\eqref{eq:g2}, we have
\begin{align}
 \langle \mathcal{G}(t)\rangle&=\, \frac{e^{S_0}}{Z(\beta)} \frac{1}{2\pi(2C)^{2\ell-2}} \frac{(2\pi)^{2\ell-1}}{\Gamma(2\ell)}\, \int d\omega e^{-i\omega t}\nonumber\\
{} & \times\textrm{exp}\left[\frac{1+\frac{1-2\ell}{\pi^2}\frac{\beta}{2C}}{\left(\frac{1}{\pi}+\frac{1-2\ell}{2\pi^3}\frac{\beta}{2C}\right)^2\frac{\beta}{2C}}\right]\textrm{exp}\left[C\omega\left(\frac{\beta}{2C}+\frac{1-2\ell}{2\pi^2}\left(\frac{\beta}{2C}\right)^2\right)\right]\times \left[\frac{\beta}{2C}+\frac{1-2\ell}{2\pi^2} \left(\frac{\beta}{2C}\right)^2\right]^{1-2\ell}\nonumber\\
 {} & \times \Gamma \left(\ell+i\frac{C\omega}{\pi} \left[\frac{\beta}{2C} + \frac{(1-2\ell)}{2\pi^2} \left(\frac{\beta}{2C}\right)^2\right]\right)\Gamma \left(\ell-i\frac{C\omega}{\pi} \left[\frac{\beta}{2C} + \frac{(1-2\ell)}{2\pi^2} \left(\frac{\beta}{2C}\right)^2\right]\right)\nonumber\\
 \approx\, & \frac{1}{e^{\frac{2\pi^2C}{\beta}}}\frac{1}{\pi^{3/2}}\left(\frac{\beta}{2C}\right)^{\frac{3}{2}}  \frac{1}{2\pi(2C)^{2\ell-2}} \frac{(2\pi)^{2\ell-1}}{\Gamma(2\ell)}\, \int d\omega e^{-i\omega t}e^{\frac{2\pi^2C}{\beta}}\nonumber\\
 {} & \times\textrm{exp}\left[C\omega\left(\frac{\beta}{2C}+\frac{1-2\ell}{2\pi^2}\left(\frac{\beta}{2C}\right)^2\right)\right]\times \left[\frac{\beta}{2C}+\frac{1-2\ell}{2\pi^2} \left(\frac{\beta}{2C}\right)^2\right]^{1-2\ell}\times\frac{\sqrt{8C\pi^3}}{\beta^{3/2}}\nonumber\\
 {} & \times \Gamma \left(\ell+i\frac{C\omega}{\pi} \left[\frac{\beta}{2C} + \frac{(1-2\ell)}{2\pi^2} \left(\frac{\beta}{2C}\right)^2\right]\right)\Gamma \left(\ell-i\frac{C\omega}{\pi} \left[\frac{\beta}{2C} + \frac{(1-2\ell)}{2\pi^2} \left(\frac{\beta}{2C}\right)^2\right]\right) \nonumber\\
  =\, &  \frac{1}{2\pi(2C)^{2\ell-2}C}\frac{(2\pi)^{2\ell-1}}{\Gamma(2\ell)}\, \int d\omega e^{-i\omega t}\textrm{exp}\left[C\omega\left(\frac{\beta}{2C}+\frac{1-2\ell}{2\pi^2}\left(\frac{\beta}{2C}\right)^2\right)\right]\left[\frac{\beta}{2C}+\frac{1-2\ell}{2\pi^2} \left(\frac{\beta}{2C}\right)^2\right]^{1-2\ell}\nonumber\\
 {} & \times \Gamma \left(\ell+i\frac{C\omega}{\pi} \left[\frac{\beta}{2C} + \frac{(1-2\ell)}{2\pi^2} \left(\frac{\beta}{2C}\right)^2\right]\right)\Gamma \left(\ell-i\frac{C\omega}{\pi} \left[\frac{\beta}{2C} + \frac{(1-2\ell)}{2\pi^2} \left(\frac{\beta}{2C}\right)^2\right]\right) \nonumber\\
 =\, & 2 \, \int \frac{d\omega}{2\pi} e^{-i\omega t}\textrm{exp}\left[\frac{\omega}{2}\times2C\left(\frac{\beta}{2C}+\frac{1-2\ell}{2\pi^2}\left(\frac{\beta}{2C}\right)^2\right)\right]\times \left[\frac{2\pi}{2C\left[\frac{\beta}{2C}+\frac{1-2\ell}{2\pi^2} \left(\frac{\beta}{2C}\right)^2\right]}\right]^{2\ell-1}\nonumber\\
 {} & \times \frac{\Gamma \left(\ell+i\frac{\omega}{2\pi}\times2C \left[\frac{\beta}{2C} + \frac{(1-2\ell)}{2\pi^2} \left(\frac{\beta}{2C}\right)^2\right]\right)\Gamma \left(\ell-i\frac{\omega}{2\pi}\times2C \left[\frac{\beta}{2C} + \frac{(1-2\ell)}{2\pi^2} \left(\frac{\beta}{2C}\right)^2\right]\right)}{\Gamma(2\ell)}\nonumber\\
 =\, & 2\,\left[\frac{\pi}{2C\left(\frac{\beta}{2C}+\frac{1-2\ell}{2\pi^2}\left(\frac{\beta}{2C}\right)^2\right){\rm sinh}\left[\frac{\pi}{2C\left(\frac{\beta}{2C}+\frac{1-2\ell}{2\pi^2}\left(\frac{\beta}{2C}\right)^2\right)}t\right]}\right]^{2\ell}\,,\label{app eq:c1}
\end{align}
where in the last equality, we have used the integral method from \cite{Lam:2018pvp}. We perform a Fourier transformation of \eqref{app eq:c1} and obtain the Wightman function
\begin{align}
\langle\mathcal{G}(\omega)\rangle &= \frac{1}{\pi}\textrm{exp}\left[\frac{\omega}{2}\times2C\left(\frac{\beta}{2C}+\frac{1-2\ell}{2\pi^2}\left(\frac{\beta}{2C}\right)^2\right)\right]\times \left[\frac{2\pi}{2C\left[\frac{\beta}{2C}+\frac{1-2\ell}{2\pi^2} \left(\frac{\beta}{2C}\right)^2\right]}\right]^{2\ell-1}\nonumber\\
 {} & \times \frac{\Gamma \left(\ell+i\frac{\omega}{2\pi}\times2C \left[\frac{\beta}{2C} + \frac{(1-2\ell)}{2\pi^2} \left(\frac{\beta}{2C}\right)^2\right]\right)\Gamma \left(\ell-i\frac{\omega}{2\pi}\times2C \left[\frac{\beta}{2C} + \frac{(1-2\ell)}{2\pi^2} \left(\frac{\beta}{2C}\right)^2\right]\right)}{\Gamma(2\ell)}\nonumber\\
 &= \frac{1}{{\rm sin}\left(\pi\ell + \pi i\frac{\omega}{2\pi}\times2C \left[\frac{\beta}{2C} + \frac{(1-2\ell)}{2\pi^2} \left(\frac{\beta}{2C}\right)^2\right] \right)}\textrm{exp}\left[\frac{\omega}{2}\times2C\left(\frac{\beta}{2C}+\frac{1-2\ell}{2\pi^2}\left(\frac{\beta}{2C}\right)^2\right)\right]\nonumber\\
 &\times\left[\frac{2\pi}{2C\left[\frac{\beta}{2C}+\frac{1-2\ell}{2\pi^2} \left(\frac{\beta}{2C}\right)^2\right]}\right]^{2\ell-1}\frac{\Gamma \left(\ell-i\frac{\omega}{2\pi}\times2C \left[\frac{\beta}{2C} + \frac{(1-2\ell)}{2\pi^2} \left(\frac{\beta}{2C}\right)^2\right]\right)}{\Gamma \left(1-\ell-i\frac{\omega}{2\pi}\times2C \left[\frac{\beta}{2C} + \frac{(1-2\ell)}{2\pi^2} \left(\frac{\beta}{2C}\right)^2\right]\right)}\, ,
\end{align}
where in the last equality we have used the identity
\begin{align}
&\Gamma \left(1-\ell-i\frac{\omega}{2\pi}\times2C \left[\frac{\beta}{2C} + \frac{(1-2\ell)}{2\pi^2} \left(\frac{\beta}{2C}\right)^2\right]\right)\Gamma \left(\ell+i\frac{\omega}{2\pi}\times2C \left[\frac{\beta}{2C} + \frac{(1-2\ell)}{2\pi^2} \left(\frac{\beta}{2C}\right)^2\right]\right)\nonumber\\
&=\frac{\pi}{{\rm sin}\left(\pi\ell + \pi i\frac{\omega}{2\pi}\times2C \left[\frac{\beta}{2C} + \frac{(1-2\ell)}{2\pi^2} \left(\frac{\beta}{2C}\right)^2\right] \right)}\,.
\end{align}
To get the imaginary part of the retarded Green's function $\langle\mathcal{G}_R(\omega,\beta)\rangle$, we use the KMS condition 
\begin{align}
&{\rm Im}\;\langle\mathcal{G}_R(\omega,\beta)\rangle = \frac{1}{2}\left(1-e^{-\beta\omega}\right)\langle\mathcal{G}(\omega)\rangle\nonumber\\
&=\frac{1}{2}\left(1-e^{-\beta\omega}\right)\frac{1}{\pi}\frac{\pi}{{\rm sin}\left(\pi\ell + \pi i\frac{\omega}{2\pi}\times2C \left[\frac{\beta}{2C} + \frac{(1-2\ell)}{2\pi^2} \left(\frac{\beta}{2C}\right)^2\right] \right)}\textrm{exp}\left[\frac{\omega}{2}\times2C\left(\frac{\beta}{2C}+\frac{1-2\ell}{2\pi^2}\left(\frac{\beta}{2C}\right)^2\right)\right]\nonumber\\
 &\times\left[\frac{2\pi}{2C\left[\frac{\beta}{2C}+\frac{1-2\ell}{2\pi^2} \left(\frac{\beta}{2C}\right)^2\right]}\right]^{2\ell-1}\frac{\Gamma \left(\ell-i\frac{\omega}{2\pi}\times2C \left[\frac{\beta}{2C} + \frac{(1-2\ell)}{2\pi^2} \left(\frac{\beta}{2C}\right)^2\right]\right)}{\Gamma \left(1-\ell-i\frac{\omega}{2\pi}\times2C \left[\frac{\beta}{2C} + \frac{(1-2\ell)}{2\pi^2} \left(\frac{\beta}{2C}\right)^2\right]\right)}\,.
\end{align}
Since for positive integer $\ell$, the retarded Green's function should be purely imaginary, one is free to write the total retarded Green's function as 
\begin{align}
\langle\mathcal{G}_R(\omega,\beta)\rangle &=i\frac{1}{2}\left(1-e^{-\beta\omega}\right)\frac{\textrm{exp}\left[\frac{\omega}{2}\times2C\left(\frac{\beta}{2C}+\frac{1-2\ell}{2\pi^2}\left(\frac{\beta}{2C}\right)^2\right)\right]}{{\rm sin}\left(\pi\ell + \pi i\frac{\omega}{2\pi}\times2C \left[\frac{\beta}{2C} + \frac{(1-2\ell)}{2\pi^2} \left(\frac{\beta}{2C}\right)^2\right] \right)}\nonumber\\
 &\times\left[\frac{2\pi}{2C\left[\frac{\beta}{2C}+\frac{1-2\ell}{2\pi^2} \left(\frac{\beta}{2C}\right)^2\right]}\right]^{2\ell-1}\frac{\Gamma \left(\ell-i\frac{\omega}{2\pi}\times2C \left[\frac{\beta}{2C} + \frac{(1-2\ell)}{2\pi^2} \left(\frac{\beta}{2C}\right)^2\right]\right)}{\Gamma \left(1-\ell-i\frac{\omega}{2\pi}\times2C \left[\frac{\beta}{2C} + \frac{(1-2\ell)}{2\pi^2} \left(\frac{\beta}{2C}\right)^2\right]\right)}\, .
\end{align}

\subsection{Quantum Green's function for $T\ll\frac{1}{C}$}\label{Sec:AppendixA2}

The $M$-dependent terms in Eq.~\eqref{eq:b1} can be rewritten as a Gaussian integral,
\begin{equation}
\int dM\; 2CMe^{\pi\frac{\omega}{2}\sqrt{\frac{2C}{M}}}e^{-M\beta}\times  \Gamma \left(\ell+i\left(\frac{\omega}{2}\sqrt{\frac{2C}{M}}\right)\right)\, \Gamma \left(\ell-i\left(\frac{\omega}{2}\sqrt{\frac{2C}{M}}\right)\right)=\int dM e^{f(M)}\,,
\end{equation}
where
\begin{equation}
f(M) = \ln{2CM}+\pi\frac{\omega}{2}\sqrt{\frac{2C}{M}}-M\beta+{\rm ln}\Gamma\left(\ell+i\frac{\omega}{2}\sqrt{\frac{2C}{M}}\right) + {\rm ln}\Gamma\left(\ell-i\frac{\omega}{2}\sqrt{\frac{2C}{M}}\right) \, .
\end{equation}
The saddle points are the solutions to the following equation,
\begin{align}
0 = \frac{\partial f(M)}{\partial M} & = \frac{1}{M}-\pi\frac{\omega}{4}\sqrt{2C}M^{-3/2}-\beta-i\frac{\sqrt{2C}\omega}{4}M^{-3/2}\psi \left(\ell + i\frac{\sqrt C\omega}{\sqrt{M}} \right)\nonumber\\
&\quad + i\frac{\sqrt{2C}\omega}{4}M^{-3/2}\psi\left(\ell-i\frac{C\omega}{\sqrt{M}}\right)\, .
\end{align}
For $\omega\ll M$ the leading-order saddle-point solutions are given by
\begin{equation}
0 = \frac{1}{M}-\beta \Rightarrow M=\frac{1}{\beta} \, .
\end{equation}
Plugging this back into Eq.~\eqref{eq:b1}, we obtain
\begin{align}
{} & \langle \mathcal{G}(t)\rangle\nonumber\\
 =&\frac{e^{S_0}}{e^{S_0+\frac{2\pi^2 C}{\beta}+\frac{3}{2}{\rm log}\frac{2\pi C}{\beta}}} \frac{4}{(2C)^{2\ell-2}}\frac{(\ell-1)!^2}{\Gamma(2\ell)}\int d\omega e^{-i\omega t}\int dM\; 2CMe^{\pi\frac{\omega}{2}\sqrt{\frac{2C}{M}}}e^{-M\beta}\nonumber\\
 {}&\times  \Gamma \left(\ell+i\left(\frac{\omega}{2}\sqrt{\frac{2C}{M}}\right)\right)\, \Gamma \left(\ell-i\left(\frac{\omega}{2}\sqrt{\frac{2C}{M}}\right)\right)\,, \nonumber\\
 \approx& e^{-\frac{2\pi^2C}{\beta}}\frac{1}{\pi^{3/2}}\left(\frac{\beta}{2C}\right)^{\frac{3}{2}} \frac{4}{(2C)^{2\ell-2}}\frac{(\ell-1)!^2}{\Gamma(2\ell)}\int d\omega e^{-i\omega t} \left(\frac{\beta}{2C}\right)^{-1}\frac{\sqrt{2\pi}}{\beta}e^{\pi\frac{\omega}{2}\sqrt{2C\beta}}e^{-1}\nonumber\\
 {}&\times  \Gamma \left(\ell+i\left(\frac{1}{2}2C\omega\sqrt{\frac{\beta}{2C}}\right)\right)\Gamma \left(\ell-i\left(\frac{1}{2}2C\omega\sqrt{\frac{\beta}{2C}}\right)\right)\,, \nonumber\\
\approx& \frac{4}{(2C)^{2\ell-2}}\frac{1}{\pi^{3/2}}\frac{(\ell-1)!^2}{\Gamma(2\ell)}e^{-1}\int d\omega e^{-i\omega t} \left(\frac{\beta}{2C}\right)^{\frac{1}{2}}\frac{\sqrt{2\pi}}{\beta} e^{\pi\frac{\omega}{2}\sqrt{2C\beta}}\nonumber\\
{}&\times\Gamma \left(\ell+i\left(\frac{1}{2}2C\omega\sqrt{\frac{\beta}{2C}}\right)\right)\, \Gamma \left(\ell-i\left(\frac{1}{2}2C\omega\sqrt{\frac{\beta}{2C}}\right)\right)\,, \nonumber\\
=& \frac{(\ell-1)!^2}{2^{2\ell-9/2}e}\left(\frac{\beta}{2C}\right)^{\ell-1}\int \frac{d\omega}{2\pi} e^{-i\omega t}  \left(\frac{2\pi}{\pi\sqrt{2C\beta}}\right)^{2\ell-1}e^{\frac{\omega}{2}\pi\sqrt{2C\beta}}\nonumber\\
{}&\times\frac{\Gamma \left(\ell+i\frac{\omega}{2\pi}\pi\sqrt{2C\beta}\right)\, \Gamma \left(\ell-i\frac{\omega}{2\pi}\pi\sqrt{2C\beta}\right)}{\Gamma(2\ell)}\,, \nonumber\\
=& \frac{(\ell-1)!^2}{2^{2\ell-9/2}e}\left(\frac{\beta}{2C}\right)^{\ell-1}\left(\frac{\pi}{\pi\sqrt{2C\beta}{\rm sinh}(\frac{\pi}{\pi\sqrt{2C\beta}}t)}\right)^{2\ell}\,,\label{eq:b6} 
 \end{align}
where in the last equality, we have used the integral method from \cite{Lam:2018pvp}. We perform a Fourier transformation of \eqref{eq:b6} and obtain the Wightman function, 
\begin{align}
&\langle\mathcal{G}(\omega)\rangle \nonumber\\
&= \frac{(\ell-1)!^2}{2^{2\ell-9/2}e}\left(\frac{\beta}{2C}\right)^{\ell-1} \frac{1}{2\pi}   \left(\frac{2\pi}{\pi\sqrt{2C\beta}}\right)^{2\ell-1}e^{\frac{\omega}{2}\pi\sqrt{2C\beta}}\frac{\Gamma \left(\ell+i\frac{\omega}{2\pi}\pi\sqrt{2C\beta}\right)\, \Gamma \left(\ell-i\frac{\omega}{2\pi}\pi\sqrt{2C\beta}\right)}{\Gamma(2\ell)} \nonumber\\
& = \frac{(\ell-1)!^2}{2^{2\ell-9/2}e}\frac{1}{\Gamma(2\ell)}\left(\frac{\beta}{2C}\right)^{\ell-1} \frac{1}{2\pi}   \left(\frac{2\pi}{\pi\sqrt{2C\beta}}\right)^{2\ell-1}e^{\frac{\omega}{2}\pi\sqrt{2C\beta}}\frac{\pi}{{\rm sin}(\pi\ell + \pi i\frac{\omega}{2}\sqrt{2C\beta} )} \frac{ \Gamma \left(\ell-i\frac{\omega}{2\pi}\pi\sqrt{2C\beta}\right)}{\Gamma(1-\ell-i\frac{\omega}{2}\sqrt{2C\beta})}\,,
\end{align}
where  in the last equality we used the identity
\begin{equation}
\Gamma(1-\ell-i\frac{\omega}{2}\sqrt{2C\beta}) \Gamma(\ell+i\frac{\omega}{2}\sqrt{2C\beta})  = \frac{\pi}{{\rm sin}(\pi\ell + \pi i\frac{\omega}{2}\sqrt{2C\beta} )} \,.
\end{equation}
To obtain the imaginary part of the retarded Green's function $\langle\mathcal{G}_R(\omega,\beta)\rangle$, we use the KMS condition 
\begin{align}
&{\rm Im}\;\langle\mathcal{G}_R(\omega,\beta)\rangle = \frac{1}{2}\left(1-e^{-\beta\omega}\right)\langle\mathcal{G}(\omega)\rangle\nonumber\\
&= \frac{1}{2}\left(1-e^{-\beta\omega}\right)\frac{(\ell-1)!^2}{2^{2\ell-9/2}e}\frac{1}{\Gamma(2\ell)}\left(\frac{\beta}{2C}\right)^{\ell-1} \frac{1}{2\pi}   \left(\frac{2\pi}{\pi\sqrt{2C\beta}}\right)^{2\ell-1}\nonumber\\
&\times e^{\frac{\omega}{2}\pi\sqrt{2C\beta}}\frac{\pi}{{\rm sin}(\pi\ell + \pi i\frac{\omega}{2}\sqrt{2C\beta} )} \frac{ \Gamma \left(\ell-i\frac{\omega}{2\pi}\pi\sqrt{2C\beta}\right)}{\Gamma(1-\ell-i\frac{\omega}{2}\sqrt{2C\beta})}\, .
\end{align}
Since for positive integer $\ell$, the retarded Green's function should be purely imaginary, one is free to write the total retarded Green's function as
\begin{align}
\langle\mathcal{G}_R(\omega,\beta)\rangle &= i\frac{1}{2}\left(1-e^{-\beta\omega}\right)\frac{(\ell-1)!^2}{2^{2\ell-9/2}e}\frac{1}{\Gamma(2\ell)}\left(\frac{\beta}{2C}\right)^{\ell-1} \frac{1}{2\pi}   \left(\frac{2\pi}{\pi\sqrt{2C\beta}}\right)^{2\ell-1}\nonumber\\
&\times e^{\frac{\omega}{2}\pi\sqrt{2C\beta}}\frac{\pi}{{\rm sin}(\pi\ell + \pi i\frac{\omega}{2}\sqrt{2C\beta} )} \frac{ \Gamma \left(\ell-i\frac{\omega}{2\pi}\pi\sqrt{2C\beta}\right)}{\Gamma(1-\ell-i\frac{\omega}{2}\sqrt{2C\beta})}\,.
\end{align}

                                                                                                                                                                                                                                                                                                                                    \subsection{Modified KMS condition}\label{Sec:AppendixA3}
From the form of the Wightman function in \eqref{eq:b6}, we see that, after a Wick rotation, the period of the Green’s function is $\pi\sqrt{2C\beta}$. 
Notably, this differs from the period $\beta$ of the Green's function in  the semiclassical limit. 
From our calculation we see that  this difference arises from a transition between different saddle points. Specifically, when moving from the semiclassical regime $(CT \gg 1)$ to the quantum regime $(CT \ll 1)$, the saddle point of the quantum-corrected Green’s function changes from $M \sim T^{2}$ to $M \sim T$. 
Given that the period is modified, 
it would be natural to  
define an \emph{effective} inverse  
temperature $\beta_{\rm eff}$, 
\begin{equation}
\beta_{\rm eff} = \pi\sqrt{2C\beta}\,,
\end{equation}
which could emerge due to nontrivial IR dynamics in the quantum phase.

Since the KMS condition (which relates the retarded and Wightman Green's functions) is based on the \emph{period} of the Wightman Green's function, 
one might be led to adopt the following modified KMS condition, 
\begin{equation}
{\rm Im}\;\langle\mathcal{G}_R(\omega,\beta)\rangle = \frac{1}{2}\left(1-e^{-\beta_{\rm eff}\omega}\right)\langle\mathcal{G}(\omega)\rangle \, 
\label{modKMS}.
\end{equation}
Using \eqref{modKMS}, instead of the standard prescription in terms of $\beta$, would then yield the following retarded Green's function:
\begin{align}
&{\rm Im}\;\langle\mathcal{G}_R(\omega,\beta)\rangle = \frac{1}{2}\left(1-e^{-\beta_{\rm eff}\omega}\right)\langle\mathcal{G}(\omega)\rangle\nonumber\\
&= \frac{1}{2}\left(1-e^{-\beta_{\rm eff}\omega}\right)\frac{(\ell-1)!^2}{2^{2\ell-9/2}e}\frac{1}{\Gamma(2\ell)}\left(\frac{\beta}{2C}\right)^{\ell-1} \frac{1}{2\pi}   \left(\frac{2\pi}{\pi\sqrt{2C\beta}}\right)^{2\ell-1}\nonumber\\
&\times e^{\frac{\omega}{2}\pi\sqrt{2C\beta}}\frac{\pi}{{\rm sin}(\pi\ell + \pi i\frac{\omega}{2}\sqrt{2C\beta} )} \frac{ \Gamma \left(\ell-i\frac{\omega}{2\pi}\pi\sqrt{2C\beta}\right)}{\Gamma(1-\ell-i\frac{\omega}{2}\sqrt{2C\beta})}\,.
\end{align}
Taking $\ell=1$ and working in the $\omega\ll T\ll C^{-1}$ limit, the retarded Green's function above can be approximated as follows,
\begin{align}
{\rm Im}\langle\mathcal{G}_R(\omega,\beta)\rangle &= \frac{2^{3/2}}{e}  e^{\frac{\omega}{2}\pi\sqrt{2C\beta}}\frac{\pi}{{\rm sinh}(\pi\frac{\omega}{2}\sqrt{2C\beta} )}\left(\frac{\omega}{2\pi}\right)\left(1-e^{-\beta_{\rm eff}\omega}\right) \nonumber \\
& \approx  \frac{2^{3/2}}{e}\omega\,.
\end{align}
Finally, the corresponding shear viscosity in the quantum regime $T\ll\frac{1}{C}$ would be
\begin{equation}
\eta' = \frac{1}{2\kappa^2}\left(\frac{r_h}{L}\right)^2\times\frac{2^{3/2}}{e}\,,
\end{equation}
which is a constant and therefore doesn't exhibit temperature dependence in the quantum regime.
We note that this result would not match the absorption cross sections of 
\cite{Emparan:2025sao,Biggs:2025nzs},
which were derived assuming the standard KMS condition.

It would be interesting to have a better understanding\footnote{We are grateful to Elias Kiritsis, Hong Liu and Sameer Murthy for comments on the KMS condition in this case.} of the physical origin of the effective inverse temperature $\beta_{\rm eff}$. This may imply the existence of a transition in the dominant spacetime geometry, from the semi-classical regime to the fully quantum regime\footnote{Similarly, to resolve the negative entropy issue in the quantum limit, one should take into account the contribution of replica wormholes; see \cite{Chandrasekaran:2022asa}.}. 
It may also arise from the approximations used to obtain \eqref{eq:b6}, which might not reflect features visible in the  full integral.
However, due to the limited understanding of hydrodynamic calculations in the quantum regime, and in order to reproduce the previous results for the cross section, in the main text we have adopted the standard KMS prescription with $\beta$. 
We look forward to having a clearer interpretation of $\beta_{\mathrm{eff}}$.

\bibliography{SV.bib}
\bibliographystyle{JHEP.bst}
\end{document}